# Phenothiazine-Based Self-Assembled Monolayer with Thiophene Head Groups Minimizes Buried Interface Losses in Tin Perovskite Solar Cells

Valerio Stacchini[§ 1,3], Madineh Rastgoo[§ 2], Mantas Marčinskas[5], Chiara Frasca[1,3], Kazuki Morita[1], Lennart Frohloff[4], Antonella Treglia[6], Thomas W. Gries, Orestis Karalis[1], Vytautas Getautis[5], Florian Ruske[1], Annamaria Petrozza[6], Norbert Koch[1,4], Hannes Hempel[1], Tadas Malinauskas[5*], Antonio Abate[1,2,3*], Artem Musiienko[7*]

1. Helmholtz-Zentrum Berlin für Materialien und Energie GmbH, 12489 Berlin, Germany

2. Department of Chemical, Materials and Production Engineering, University of Naples Federico II, Fuorigrotta, Italy

3. Department of Chemistry, University of Bielefeld, Universitätsstraße 25, 33615 Bielefeld, Germany

4. Institut für Physik & Center for the Science of Materials (CSMB), Humboldt-Universität zu Berlin, 12489, Berlin

5. Department of Organic Chemistry, Kaunas University of Technology, Radvilenu pl. 19, Kaunas LT-50254, Lithuania

6. Istituto Italiano di Tecnologia (IIT), via Rubattino 81, 20134, Milan, Italy

7. Young Investigator Group, Robotized Material and Photovoltaic Engineering, Helmholtz-Zentrum Berlin für Materialien und Energie (HZB), Berlin, Germany.

§ These authors contributed equally to this work

* Corresponding authors: artem.musiienko@helmholtz-berlin.de, antonio.abate@helmholtz-berlin.de, tadas.malinauskas@ktu.lt

# Abstract

Self-assembled monolayers (SAMs) have revolutionised the fabrication of lead-based perovskite solar cells, but they still remain underexplored in tin perovskite systems. To date, PEDOT remains the most effective hole-selective layer in tin perovskite solar cells (TPSCs), yet it presents challenges for both performance and stability. MeO-2PACz, the only SAM reported for tin perovskites consistently underperforms when compared to PEDOT. In this work, we identify that MeO-2PACz's limitations stem from excessively strong interactions with the perovskite surface and poor lattice matching, which leads to inferior interface quality. To address these issues, we design, synthesise, and characterise a novel SAM-forming molecule called Th-2EPT. We used density functional theory (DFT) to evaluate coordination strength and lattice compatibility, complemented by electro-optical characterisation techniques that show significantly reduced interfacial recombination and improve material crystallinity in Th-2EPT/Perovskite films. With Th-2EPT, we demonstrated the first SAM-based tin perovskite solar cells that outperform PEDOT-based devices, delivering a power conversion efficiency (PCE) of 8.2% with a DMSO-free solvent system.

# Introduction

Thanks to their simple processing, bandgap tunability, and defect tolerance, perovskites are on track to radically change the world of photovoltaics. However, two challenges must be addressed before this technology can be considered competitive for commercialisation: long-term stability and lead toxicity.[1–4] Intense research is focused on developing non-toxic alternatives to mitigate the toxicity issues of lead-containing perovskites.[2,5,6]. Tin perovskite solar cells are potentially even more stable than their lead-based counterparts, exhibiting negligible ion migration.[3,7,8] Despite their potential, TPSCs underperform compared to their lead counterparts, hindered by rapid crystallisation and unstable interfaces.[2], due to the facile oxidation of $Sn^{2+}$ to $Sn^{4+}$ and poor energy-level alignment between standard selective contacts and the absorber layer.[1,2,8–10]

Until now, in most studies on tin perovskite solar cells, poly(3,4-ethylene dioxythiophene) polystyrene sulfonate, in short PEDOT: PSS, serves as the gold-standard hole-selective layer (HSL) in current tin perovskite solar cell. However, the significant energetic misalignment, acidity, and hygroscopicity limit the performance of PEDOT-based devices and lead to device degradation.[2,11–14] Due to the inherent chemical limitations of PEDOT, overcoming its drawbacks calls for the design of novel materials. Self-assembled monolayers (SAMs) are emerging as a key alternative to PEDOT due to their dual role in interface passivation and charge selectivity.[15–17] While SAMs have demonstrated success in lead-based perovskites by improving charge extraction and passivating interfaces, their application in tin perovskites remains underexplored. To the best of our knowledge, MeO-2PACz is the only SAM successfully employed for tin perovskites, achieving moderate power conversion efficiencies (PCEs) of 5.8%[9] and 9.4%,[18] obtained with a DMSO-free and DMSO solvent system, respectively. Regardless of the solvent system, SAM-based solar cells have yet to outperform PEDOT controls, and the underlying causes of poor performance remain unclear.[2,19–21] In this work, we use first-principles calculations to identify the limitations of MeO-2PACz, revealing its strong interaction with the $FASnI_3$ lattice and dimensional mismatch as the cause for a highly defective interface. We employ a DMSO-free solvent system for tin PSCs to mitigate solvent-induced degradation, addressing two key challenges: the oxidation of $Sn^{2+}$ by DMSO and the presence of residual DMSO, which can further promote oxidation over the device's lifetime.

To overcome this incompatibility between the MeO-2PACz and tin perovskite, we design and synthesise a novel SAM molecule, {2-[3,7-di(thiophen-3-yl)-10*H*-phenothiazin-10-yl]ethyl}phosphonic acid or Th-2EPT, featuring two thiophene passivating head groups, a phenothiazine core and a phosphonic acid anchoring group. Th-2EPT significantly improved the interfacial quality of tin perovskites by enhancing lattice matching and fine-tuning interaction strength. . This work marks the first successful attempt to fabricate well-functioning SAM-based tin perovskite solar cells. Th-2EPT is able to compete with and outperform PEDOT, thanks to positive chemical interaction, lattice matching and improved optoelectronic properties of perovskite films grown on this SAM.

# Results and Discussion

To investigate the underperformance of MeO-2PACz in tin perovskite solar cells, we examined its interfacial interaction with the perovskite lattice using density functional theory (DFT). Specifically, we evaluated two key descriptors of interface quality: the binding energy of the SAM to the $FASnI_3$'s uncoordinated $Sn^{2+}$ surface ions  and the degree of lattice matching between the SAM's passivating head groups and the perovskite cationic sublattice.

In the case of MeO-2PACz, we simulated the molecule without its phosphonic acid anchor to isolate the interaction of the two methoxy head groups with undercoordinated $Sn^{2+}$ ions. The molecule was fully relaxed on the slab model, and the binding energy was calculated as the interaction energy between the donor atoms and the surface cations (full computational details are provided in the Supporting Information). MeO-2PACz exhibited a high binding energy of 0.45 eV (Figure 1c), in line with strong Lewis acid–base adduct formation between its electronegative oxygen atoms and $Sn^{2+}$.

While this strong interaction could be beneficial for anchoring, it may also rigidly lock the interface, preventing the perovskite lattice from undergoing necessary rearrangements during film formation. Previous studies have suggested that strong molecular interactions can sometimes be detrimental, as they may rigidly anchor the interface, hindering the natural rearrangements of the perovskite lattice during crystal growth.[17,22,23]

Next, we evaluated the spatial compatibility between the molecule's O–O spacing (7.81 Å) and the calculated Sn–Sn distances on the perovskite surface (7.81 Å - Figure 1a). The best match was found with the second-nearest neighbours (9.22 Å), corresponding to a rather low 85% match (Figure 1b). A detailed calculation on lattice matching is discussed in the Supporting Information. We hypothesize that, in the case of MeO-2PACz, the combination of strong interactions with undercoordinated $Sn^{2+}$ ions and poor lattice compatibility may induce interfacial strain and promote defect formation.

To address lattice incompatibility of MeO-2PACz, we designed a new SAM molecule, Th-2EPT, with two main objectives: reduce binding strength and improve lattice compatibility. In place of the carbazole core used in MeO-2PACz, we introduced a phenothiazine unit - a well-established, low-cost electron-rich structure.[24–26] To reduce the binding strength observed with methoxy groups, we replaced them with thiophene units, which were introduced at positions 3 and 7 of the phenothiazine core (Figure 1d). Thiophenes were selected due to their lower electronegativity relative to oxygen, while still offering proven effectiveness in surface passivation for both tin- and lead-based perovskites[27,28]. We retained the phosphonic acid anchor, given its proven effectiveness in ensuring robust adhesion[29,30]. The resulting molecular structure, visible in Figure 1d, named diethyl{2-[3,7-

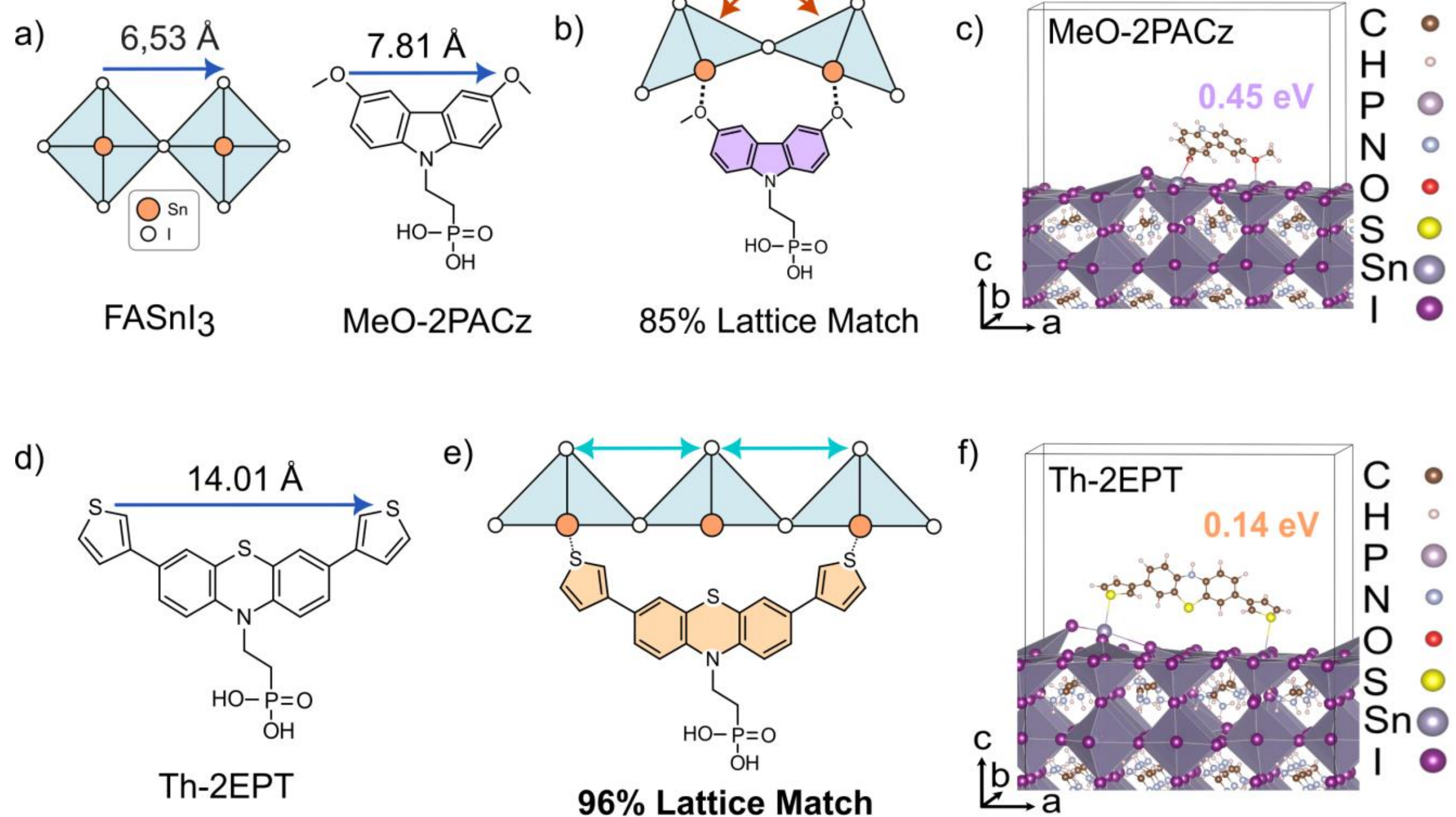

**Figure 1**: **SAM-Perovskite Interaction and Lattice Match. (a)** Structures and characteristic dimensions of FASnI3 lattice and the MeO-2PACz phosphonic acid. **(b)** Graphical representation of lattice mismatch between MeO-2PACz and the FASnI3 lattice. **(c,f)** DFT-calculated binding energies for the two molecules on top of a $SnI_2$-terminated $FASnI_3$ surface. The values in eV denote the computed binding energies at the interface. **(d)** Molecular structure of Th-2EPT and S-S distance between head groups. **(e)** Lattice matching between Th-2EPT and the FASnI lattice. For graphical clarity, the illustration simplifies the 96% lattice match of Th-2EPT by representing it as alignment with two lattice parameters; the full calculation with fifth-nearest-neighbour matching is provided in the Supporting Information.

di(thiophen-3-yl)-10H-phenothiazin-10-yl]ethyl}phosphonate (Th-2EPT), was investigated with the same DFT methodology. Th-2EPT exhibited a substantially lower binding energy of 0.14 eV (Figure 1f), attributed to the less electronegative sulfur atoms and their more delocalized bonding character. This weaker interaction suggests a more flexible, less disruptive binding mode that could facilitate perovskite crystallization. Geometrically, Th-2EPT showed a near-ideal S–S spacing of ~14.01 Å, matching the fifth-nearest Sn–Sn distance in $FASnI_3$ (~14.6 Å) with 96% alignment (for detailed calculations, see Supporting Information). The superior geometric match, combined with moderate binding strength, indicates that Th-2EPT could passivate the perovskite surface without introducing strain or defects.

Motivated by the favourable binding energy and lattice compatibility predicted for Th-2EPT, we synthesized the molecule and experimentally tested its effect on $FASnI_3$ film formation (synthetic steps are summarized in Table 1 of the Supporting Information, followed by a detailed description of the synthetic intermediates). We deposited $FASnI_3$ perovskite on PEDOT, MeO-2PACz and Th-2EPT on ITO substrates. PEDOT was included as a reference in this stage of the study, given its role as the most well-known hole transport layer (HTL) in $FASnI_3$ solar cells.[9,21] PEDOT and MeO-2PACz were spin-coated, while target Th-2EPT was deposited through dip-coating. In the case of PEDOT films, an alumina ($Al_2O_3$) nanoparticles scaffold is spin-coated between PEDOT and perovskite. This step is necessary to improve the wetting of PEDOT.[21] Contact angle measurements (Figure 2a-c) confirmed good wettability for Th-2EPT (69°) and good wettability for PEDOT (59° with the spin-coated alumina nanoparticle wetting layer), and MeO-2PACz (55°). The resulting films were characterized by SEM and XRD to study the film morphology and effect of new Th-2EPT SAM on crystallization (Figure 2d,e). XRD shows reflexes from both the ITO substrate and the perovskite layer,[31] and all peaks could be assigned to either of the two materials. XRD diffractograms for the three samples do not exhibit significant differences in the relative intensities of diffraction peaks, suggesting comparable crystallographic orientation across all HTL layers. However, notable differences are observed in the analysis of the primary (100) diffraction peak located around 14° (Figure 2e). The intensity of this peak is similar for PEDOT and Th-2EPT, whereas it is markedly reduced for MeO-2PACz, indicating lower crystallinity. Peak broadening was evaluated using a Gaussian fitting model, and the full width at half maximum (FWHM) was extracted (more information on the fitting model in the supporting information). The extracted FWHM show a similar trend with respect to the same signal's intensity, with MeO-2PACz showing by far the broadest peak ($FWHM_{MeO}$ = 0.30°) while PEDOT shows the narrowest ($FWHM_{PEDOT}$ = 0.10°), followed by Th-2EPT ($FWHM_{Th\text{-}2EPT}$ = 0.12°). The pronounced peak broadening observed for the film on MeO-2PACz may arise from several factors, including increased microstrain, a higher density of structural defects, reduced crystallinity, or a smaller vertical grain size. Regardless of the dominant mechanism, the broadening indicates a lower structural quality for MeO-2PACz films relative to the films deposited on PEDOT or Th-2EPT. While displaying a very similar FWHM, films on PEDOT and Th-2EPT show a rather different surface morphology in SEM (Figure 2a,c). Films grown on PEDOT exhibit the largest lateral grain size, whereas those deposited on both SAMs display notably smaller and more compact grains. Likely, the grain growth is mainly governed by the usage of an alumina scaffold in the case of PEDOT.

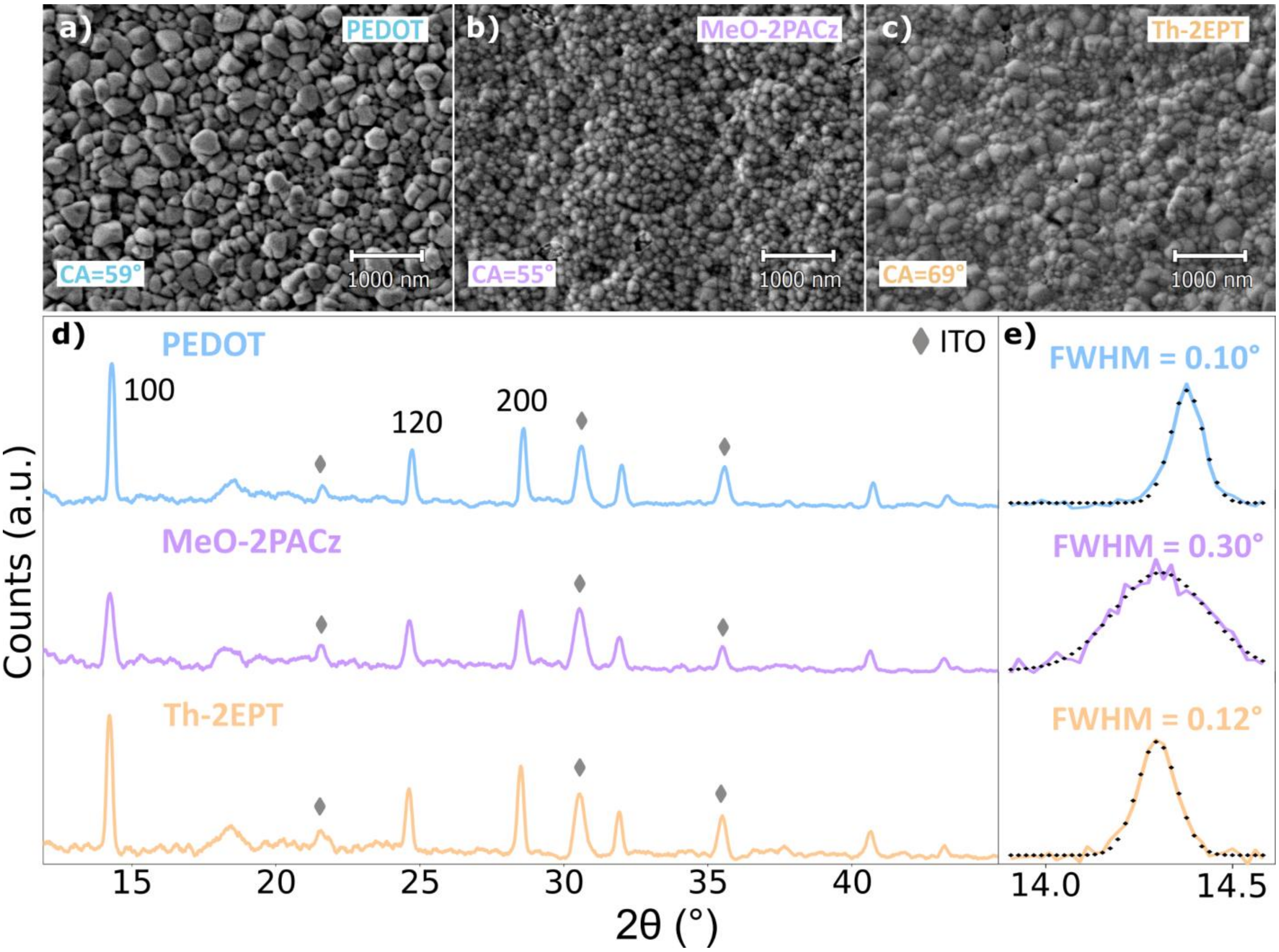

**Figure 2: Morphology and Crystallinity of Perovskite Films.** SEM images of perovskite films deposited on PEDOT **(a)**, MeO-2PACz **(b)**, and Th-2EPT **(c)**. Median values for measured contact angles (CA) are indicated (CA data is provided in Figure S8, Supporting Information). **(d)** X-ray diffraction (XRD) patterns of the same films (left), with principal diffraction peaks indexed and ITO substrate peaks marked by diamonds. **(e)** Magnified view of the (100) diffraction peak around 14°, showing fitted curves (black diamonds) and corresponding full width at half maximum (FWHM) values for each sample, extracted using a gaussian fitting. More information on the fitting is provided in the Supporting Information.

After observing that Th-2EPT significantly improves film crystallinity and morphology - surpassing MeO-2PACz and closely approaching PEDOT - we next investigated how these structural enhancements affect the optoelectronic properties and carrier recombination dynamics of the films. ITO/HTL/Perovskite layer stacks were studied, always illuminating them from the ITO/HTL side, to better probe the buried interface (Figure 3a). We employed PLQY, trPL, and TAS measurements to assess how the hole transport layer (HTL) influences radiative efficiency and carrier recombination in perovskite films (Figure 3). Figure 3b presents the normalised photoluminescence quantum yield (PLQY) as a function of excitation density for perovskite films deposited on various substrates. At low excitation densities ($10^{14}$ $cm^{-3}$), the film on Th-2EPT exhibits the highest PLQY, surpassing even that of the reference glass substrate. This elevated PLQY at low carrier concentrations indicates effective suppression of trap-assisted non-radiative recombination, suggesting superior interfacial passivation provided by Th-2EPT. As excitation density increases ($10^{16}$-$10^{17}$ $cm^{-3}$), the PLQY values for Th-2EPT, MeO-2PACz, and glass converge, forming a group with similar peak efficiencies. In contrast, the PEDOT-based film consistently displays significantly lower PLQY across the entire excitation range, indicative of higher non-radiative recombination losses and inferior interfacial quality. These observations underscore the efficacy of SAM-based hole transport layers in enhancing radiative recombination efficiency and mitigating interfacial defects. Transient photoluminescence (trPL) was measured on ITO/HTL/Perovskite stacks. The photoluminescence transients reveal clear differences in decay times among the hole transport layers, with Th-2EPT exhibiting the slowest decay (Figure 3c). The trPL data were fitted using a bi-exponential decay model (fitting parameters and corresponding plots are provided in Figure S6 of the Supporting Information). The resulting fits were then used to compute differential lifetimes according to the expression $t = -\{d\ ln[f(t)]/dt\}^{-1}$, where $f(t)$ is the time-dependent photon flux. In a bi-exponential trPL fit, the fast-decaying exponential models the initial behaviour of the decay and is more closely associated with interfacial trap-assisted non-radiative recombination. In contrast, the slowly decaying exponential is more representative of carrier lifetimes.[32,33] Both lifetimes are tabulated in Figure S6 of the Supporting Information. While the fast components do not appear in the differential lifetime plots, the slow decay manifests as a plateau in these graphs (Figure 3d). PEDOT reaches this plateau at only 35 ns, indicating the shortest lifetime among the evaluated HTLs, followed by MeO-2PACz with a lifetime of 67 ns. The film on Th-2EPT shows a longer lifetime of 91 ns, while the perovskite grown on quartz exhibits a lifetime of 156 ns - an impressive value when compared to recent studies reporting significantly shorter trPL lifetimes for Sn-based perovskites.[34,35] Following trPL, transient absorption spectroscopy (TAS) was performed. TAS measures the relative change in transmittance (photobleach) of the material as a function of time after photoexcitation. The resulting signal is proportional to the population of holes and electrons in the valence and conduction bands, respectively. The main difference of TAS with respect to trPL is that it enables tracking of both radiative and non-radiative recombination paths, while photoluminescence techniques only tack radiative recombination. With a temporal resolution of approximately 300 fs, TAS enables the study of excited-state dynamics across a broad time range, extending up to tens of microseconds and spanning seven orders of magnitude in resolution. For this measurement, samples were illuminated from the HTL side using a 343 nm laser, selected for its shallow penetration depth in perovskite, thereby ensuring excitation is localised near the buried interface. In Figure 3e, carrier dynamics are extracted at the photobleach maximum around 850 nm (full spectra available in the Supporting Information, Figure S4). A trend consistent with trPL is observed: the perovskite film on PEDOT exhibits the fastest decay. At the same time, those on MeO-2PACz and Th-2EPT show progressively longer lifetimes, approaching the behaviour of perovskite grown on glass. Figure 3f presents the TAS decay in the microseconds range, revealing distinct decay rates for the different HTLs. The signal for Th-2EPT decays significantly faster than for PEDOT and MeO-2PACz. A rapid decay in this time window is associated with a lower density of trapped carriers, while a slower decay indicates a higher trap population. This occurs because trapped carriers sustain the TAS signal indirectly by leaving behind an uncompensated charge in the opposite band. The longer this charge imbalance persists, the longer the signal is maintained. The fast decay observed for Th-2EPT suggests a reduced density of trapped carriers, supporting its beneficial role in defect passivation and improved interaction with the perovskite. In contrast, MeO-2PACz shows a prolonged signal, consistent with a higher trap population. In summary, PEDOT-based films exhibit short carrier lifetimes and low PLQY across all excitation densities, consistent with high non-radiative losses and poor buried interface quality. MeO-2PACz shows improved optoelectronic performance relative to PEDOT, but still exhibits signs of interfacial limitations. Both PEDOT and MeO-2PACz display prolonged TAS signals (Figure 3f), indicating long-lived trap states at the interface. Th-2EPT, by contrast, shows the slowest trPL decay, the fastest TAS signal decay in the microsecond regime, and the highest PLQY at low excitation densities - even exceeding that of glass. These results point to efficient interfacial passivation and improved carrier extraction, underscoring the effectiveness of Th-2EPT as a hole-selective interface layer for tin perovskite solar cells.

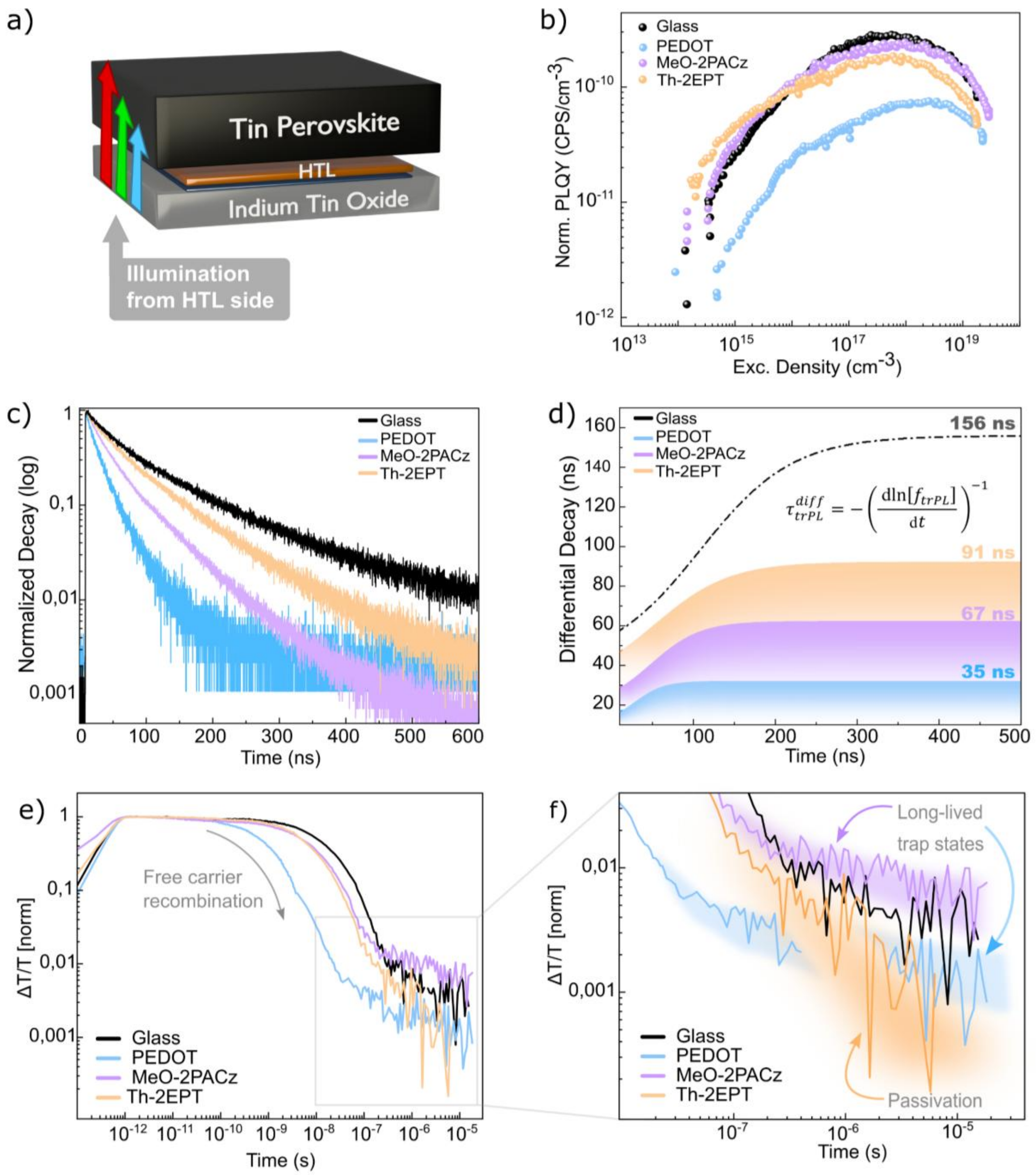

**Figure 3 Optoelectronic Properties (a)** Schematic representation of the ITO/HTL/PVK stack on which the measurements were performed. The RGB coloured arrows qualitatively represent the penetration depth dependence on the excitation wavelength. **(b)** PLQY plots were obtained by integrating the PL spectrum as a function of excitation density. **(c)** Normalised nanosecond-resolved photoluminescence transients (trPL) and **(d)** computed differential lifetimes from double exponential fits to the trPL transients. **(e)** Transient absorption (TAS) decays, showing the decay of the photo bleach peak at 850 nm, with an excitation density of $4{\cdot}10^{17}$ cm$^{-3}$. **(f)** A magnified view of the TAS decay, focusing on the microseconds time range, highlights the distinct long-timescale recombination dynamics and reveals apparent differences in decay rates among the HTL substrates. Plot e) and f) only show the decay of the 850 nm peak, while full transient absorption spectra are available in the Supporting Information, Figure S5.

To implement improvements in optoelectronic properties and assess the impact of the SAMs on device performance, we fabricated solar cells incorporating MeO-2PACz, PEDOT, and Th-2EPT as hole-selective layers and tin perovskite $EDA_{0.05}FA_{0.95}SnI_3$ as the absorber. Our approach avoids the use of DMSO, which, despite being the most commonly used solvent for preparing tin perovskite, has also been identified as one of the sources of the $Sn^{2+}$ oxidation[36,37]. Instead, we employed a solvent system composed of DEF (N, N'-Diethylformamide) and DMPU (N,N′-Dimethylpropyleneurea). Among the candidates, the phenothiazine-based Th-2EPT demonstrated superior performance, achieving a record efficiency of 8.2% and surpassing MeO-2PACz (4.5%) and PEDOT (7.1%). The device architecture consists of ITO/HSL/DMSO-free Sn Perovskite/C60/BCP/Ag, where the HSL was varied between the three materials. Spin-coating and dip-coating were evaluated for Th-2EPT, with superior results obtained via dip-coating. The solar cell performance is summarised in Figure 4, which shows champion J-V curves (Figure 4b), EQE spectra (Figure 4c), Voc boxplots (Figure 4d), and PCE boxplots (Figure 4e). Th-2EPT demonstrated improved short-circuit current density (Jsc), delivering 18.8 mA/cm² compared to MeO-2PACz (14.9 mA/cm²) and PEDOT (18.4 mA/cm²). Incident photon current efficiency (IPCE) measurements were performed on the same champion pixels from the J-V measurements to investigate this improvement. The IPCE spectrum confirmed the high Jsc of Th-2EPT devices, with a value of 18.97 mA $cm^{-2}$, compared to 14.99 mA $cm^{-2}$ for MeO-2PACz and 17.67 mA $cm^{-2}$ for PEDOT (Figure 4c). The EQE spectrum highlights the superior quantum efficiency of Th-2EPT, particularly at shorter wavelengths, with a high internal photon-to-electron conversion at 400 nm, far exceeding PEDOT and MeO-2PACz devices. Short wavelengths in the range 350-450 nm have low penetration depth in perovskite and are readily absorbed in the first nanometres, in the proximity of the HTL/Perovskite interface. For this reason, a substantial IPCE improvement at low wavelengths suggests improved interface quality and reduced non-radiative recombination at the Th-2EPT interface[38]. Open circuit voltage (Voc) is also significantly enhanced in Th-2EPT devices compared to PEDOT and MeO-2PACz. Champion Voc values of 0.63 V, 0.57 V, and 0.50 V were recorded for Th-2EPT, PEDOT, and MeO-2PACz, respectively (Figure 4d). Th-2EPT champion devices outperform PEDOT and MeO-2PACz in all solar cell parameters. Fill factor and short circuit current boxplots are shown in Figure S7 of the Supporting Information. With 8.2% PCE, our Th-2EPT champion device is a record for SAM-based, DMSO-free tin perovskite solar cells to the best of our knowledge. Figure S7 presents a comparison of short-term operational stability under maximum power point (MPP) tracking of TPSC devices with different HSLs in a nitrogen environment. To gain more information on the buried interface from solar cells, we measured Voc dependence on light intensity to extract the ideality factor, which quantifies deviation from ideal diode behaviour, and it is a valuable index of non-radiative recombination at the interface.[39] The measurement was performed on working solar cells, displaying the same architecture used for devices, as shown in Figure 4a. While an ideal cell has $n = 1$, tin perovskite solar cells can show high ideality factors due to intense bulk and interfacial recombination.[11,40] By measuring Voc across varying light intensities, the ideality factor was extracted using an equation easily derived from the diode equation in open circuit conditions:

$$n = \frac{q}{kT} \frac{dV_{oc}}{d(\ln(I))}$$

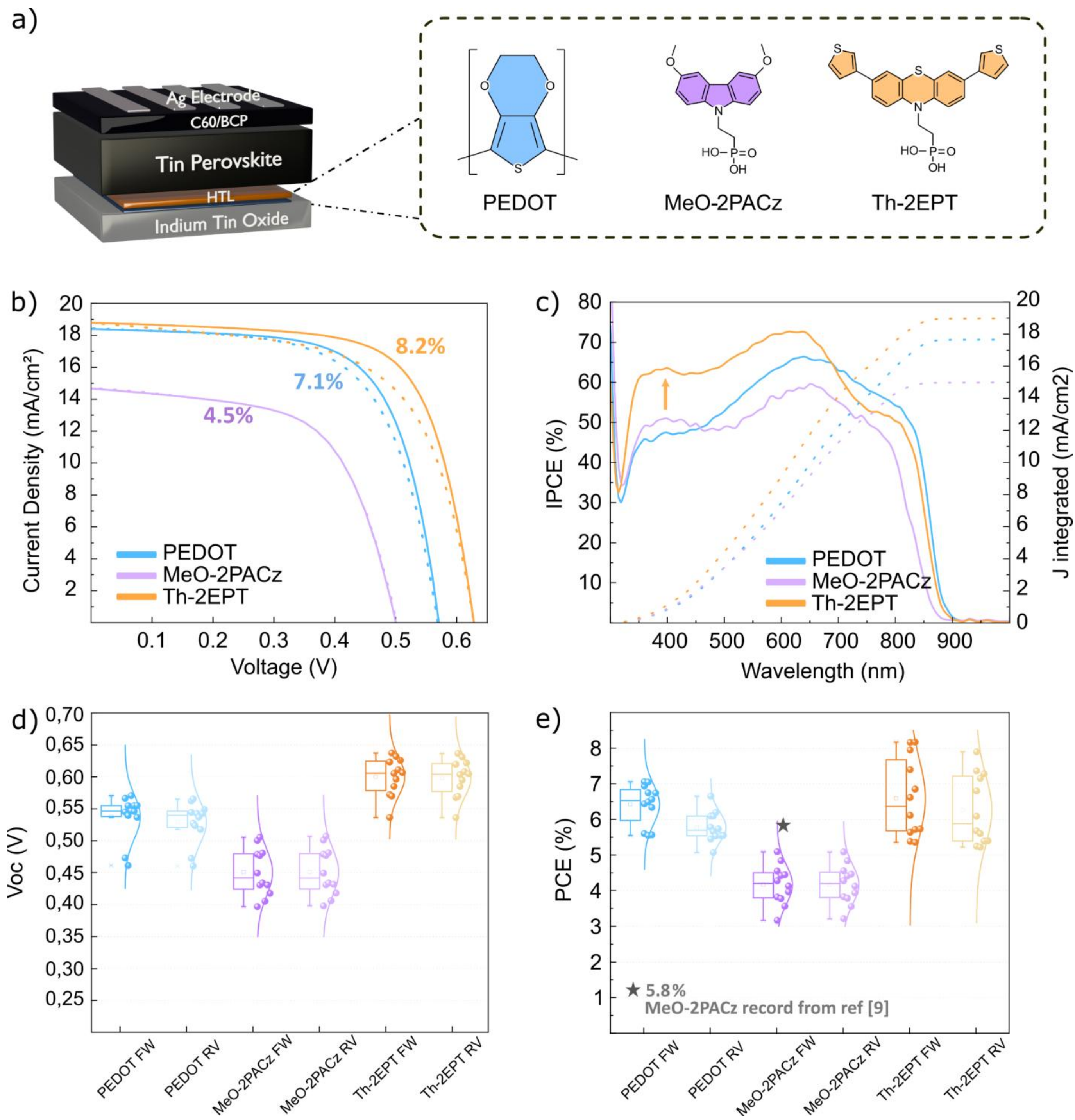

**Figure 4 Solar Cell Performance a)** Schematic device structure of the lead-free solar cell used in this work (left) and molecular structures of the three different hole-selective materials (right). **b)** J-V characteristics and **c)** EQE spectra for the champion device for each hole-selective layer. Measured pixels in b) and c) are the same. Boxplots showing statistics for **d)** Open circuit voltage ($V_{OC}$) and **e)** power conversion efficiency (PCE) for the measured solar cells. Displayed with a grey star in plot e), the highest PCE was obtained with a SAM in a DMSO-free perovskite solar cell, as shown by the work of Aktas et al.[9]

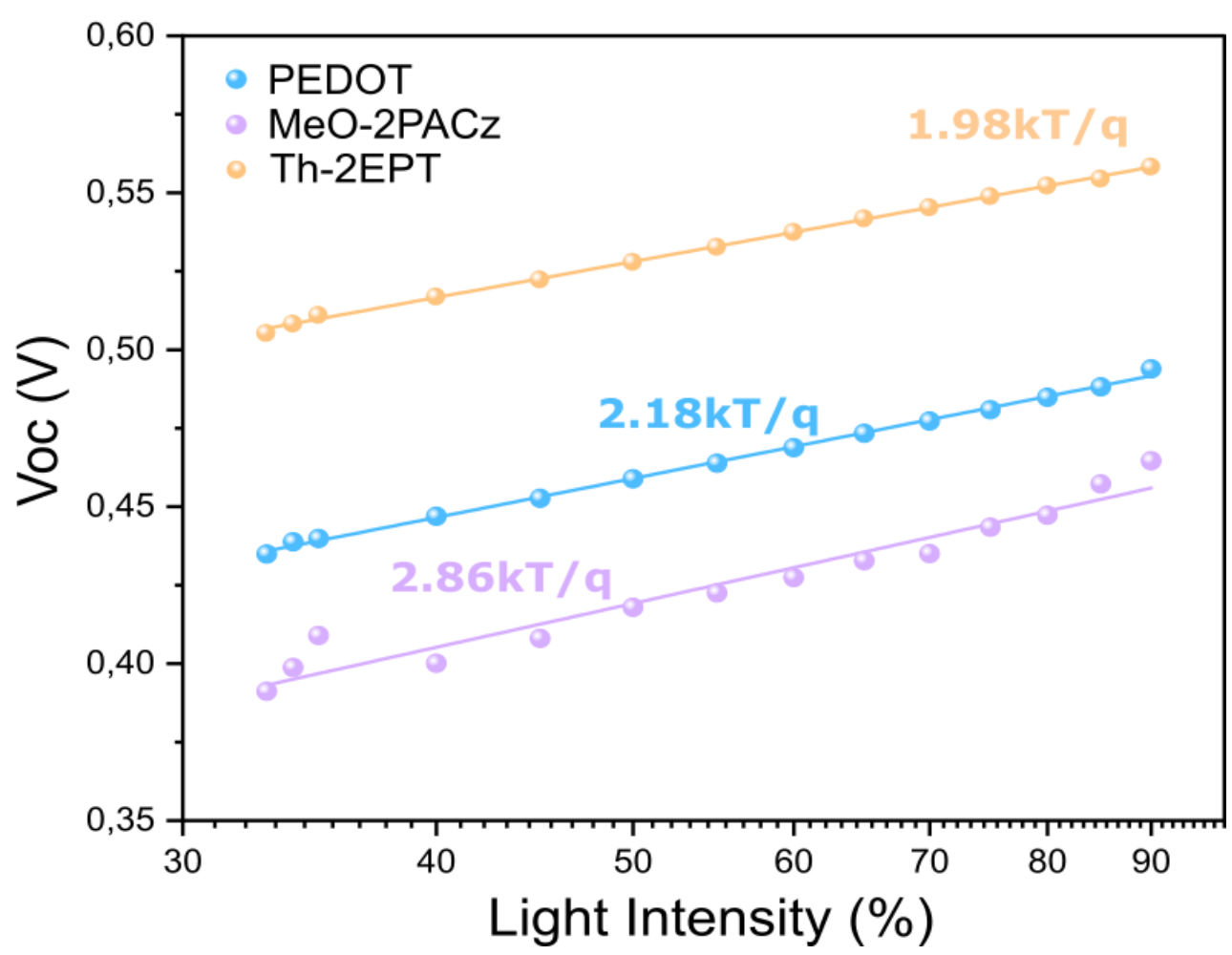

**Figure 5** Voc dependence on light intensity

The extracted values for *n* resulted in 2.86, 2.18, and 1.98 for MeO-2PACz, PEDOT, and Th-2EPT, respectively (Figure 5). The lower ideality factor with Th-2EPT is an indicator of reduced non-radiative recombination. This result confirms the ability of Th-2EPT to drastically reduce non-radiative recombination, promoting the formation of a low-defect interface through an optimized interaction.

# Conclusion

The majority of self-assembled monolayers (SAMs) employed as hole-selective layers in tin-based perovskite solar cells have failed to surpass the performance of PEDOT, despite its acidity, energy mismatch, and instability. In this work, we designed and synthesized a novel SAM molecule, Th-2EPT, specifically tailored to enhance lattice matching and introduce bond flexibility—addressing the limitations of MeO-2PACz. Compared to PEDOT, Th-2EPT enables the formation of perovskite films with comparable crystallinity, albeit with smaller grains. However, it delivers markedly superior optoelectronic performance. While PEDOT benefits from favorable energetic alignment with $FASnI_3$, it suffers from poor carrier dynamics, as evidenced by short lifetimes and low PLQY. In contrast, Th-2EPT demonstrates longer carrier lifetimes and higher photoluminescence efficiency, pointing to reduced non-radiative recombination. Importantly, Th-2EPT supports the fabrication of high-quality, DMSO-free devices and outperforms both PEDOT and MeO-2PACz in terms of optoelectronic quality and overall device performance. The results confirm that Th-2EPT forms an exceptionally well-passivated buried interface, effectively minimizing interfacial recombination losses**.** These findings establish Th-2EPT as a promising candidate for next-generation SAM-based architectures and demonstrate that rational molecular design can unlock new levels of performance in tin perovskite photovoltaics.

# Acknowledgments

This project has received funding from the German Federal Ministry of Education and Research (Bundesministerium für Bildung und Forschung, BMBF) under the NanoMatFutur Call, project number 03XP0625, COMET PV, and the European Union’s Framework Program for Research and Innovation HORIZON EUROPE (2021-2027) under the Marie Skłodowska-Curie Action Postdoctoral Fellowships (European Fellowship) 101061809 HyPerGreen. M.M., V.G. and T.M. would like to acknowledge the project “Mission-driven Implementation of Science and Innovation Programmes” (No. 02-002-P-0001), funded by the Economic Revitalization and Resilience Enhancement Plan “New Generation Lithuania”. L.F. acknowledges financial support by the German

Federal Environmental Foundation (DBU). A.T. acknowledges the Italian Ministry of Environment and Energy Security in the framework of the Project GoPV (CSEAA_00011) for Research on the Electric System. K.M. acknowledges the Gauss Centre for Supercomputing e.V. for providing computing time on the GCS Supercomputer SuperMUC-NG and the Alexander von Humboldt Foundation for the research funding.

# Data Availability

The data that support the findings of this study are available from the corresponding author upon reasonable request.

# References

1. Aldamasy, M. *et al.* Challenges in tin perovskite solar cells. *Phys. Chem. Chem. Phys.* **23**, 23413–23427 (2021).
2. Aktas, E. *et al.* Challenges and strategies toward long-term stability of lead-free tin-based perovskite solar cells. *Commun. Mater.* **3**, 104 (2022).
3. Abate, A. Stable Tin-Based Perovskite Solar Cells. *ACS Energy Lett.* **8**, 1896–1899 (2023).
4. Dey, K. *et al.* Substitution of lead with tin suppresses ionic transport in halide perovskite optoelectronics. *Energy Environ. Sci.* **17**, 760–769 (2024).
5. Babayigit, A. *et al.* Assessing the toxicity of Pb- and Sn-based perovskite solar cells in model organism Danio rerio. *Sci. Rep.* **6**, 18721 (2016).
6. Li, J. *et al.* Biological impact of lead from halide perovskites reveals the risk of introducing a safe threshold. *Nat. Commun.* **11**, 310 (2020).
7. Le, Z. *et al.* Ion Migration in Tin-Halide Perovskites. *ACS Energy Lett.* **9**, 1639–1644 (2024).
8. Li, G. *et al.* Ionic Liquid Stabilizing High-Efficiency Tin Halide Perovskite Solar Cells. *Adv. Energy Mater.* **11**, 2101539 (2021).
9. Aktas, E. *et al.* One-Step Solution Deposition of Tin-Perovskite onto a Self-Assembled Monolayer with a DMSO-Free Solvent System. *ACS Energy Lett.* **8**, 5170–5174 (2023).

10. Cui, D. *et al.* Making Room for Growing Oriented $FASnI_3$ with Large Grains via Cold Precursor Solution. *Adv. Funct. Mater.* **31**, 2100931 (2021).

11. Zhang, X. *et al.* The Voltage Loss in Tin Halide Perovskite Solar Cells: Origins and Perspectives. *Adv. Funct. Mater.* **32**, 2108832 (2022).

12. Xia, Y., Yan, G. & Lin, J. Review on Tailoring PEDOT:PSS Layer for Improved Device Stability of Perovskite Solar Cells. *Nanomaterials* **11**, 3119 (2021).

13. Di Girolamo, D. *et al.* Enabling water-free PEDOT as hole selective layer in lead-free tin perovskite solar cells. *Mater. Adv.* **3**, 9083–9089 (2022).

14. Cameron, J. & Skabara, P. J. The damaging effects of the acidity in PEDOT:PSS on semiconductor device performance and solutions based on non-acidic alternatives. *Mater. Horiz.* **7**, 1759–1772 (2020).

15. Kim, S. Y., Cho, S. J., Byeon, S. E., He, X. & Yoon, H. J. Self-Assembled Monolayers as Interface Engineering Nanomaterials in Perovskite Solar Cells. *Adv. Energy Mater.* **10**, 2002606 (2020).

16. Liu, M. *et al.* Compact Hole-Selective Self-Assembled Monolayers Enabled by Disassembling Micelles in Solution for Efficient Perovskite Solar Cells. *Adv. Mater.* **35**, 2304415 (2023).

17. Azam, M. *et al.* Dual functionality of charge extraction and interface passivation by self-assembled monolayers in perovskite solar cells. *Energy Environ. Sci.* **17**, 6974–7016 (2024).

18. Song, D., Ramakrishnan, S., Zhang, Y. & Yu, Q. Mixed Self-Assembled Monolayers for High-Photovoltage Tin Perovskite Solar Cells. *ACS Energy Lett.* **9**, 1466–1472 (2024).

19. Song, D., Narra, S., Li, M.-Y., Lin, J.-S. & Diau, E. W.-G. Interfacial Engineering with a Hole-Selective Self-Assembled Monolayer for Tin Perovskite Solar Cells via a Two-Step Fabrication. *ACS Energy Lett.* **6**, 4179–4186 (2021).

20. Fan, X. *et al.* PEDOT:PSS for Flexible and Stretchable Electronics: Modifications, Strategies, and Applications. *Adv. Sci.* **6**, 1900813 (2019).

21. Di Girolamo, D. *et al.* Enabling water-free PEDOT as hole selective layer in lead-free tin perovskite solar cells. *Mater. Adv.* **3**, 9083–9089 (2022).

22. Liu, D. *et al.* Strain analysis and engineering in halide perovskite photovoltaics. *Nat. Mater.* **20**, 1337–1346 (2021).

23. Jiang, W., Hu, Y., Li, F., Lin, F. R. & Jen, A. K.-Y. Hole-Selective Contact with Molecularly Tailorable Reactivity for Passivating High-Performing Inverted Perovskite Solar Cells. *CCS Chem.* **6**, 1654–1661 (2024).

24. Maciejczyk, M. R., Chen, R., Brown, A., Zheng, N. & Robertson, N. Beyond efficiency: phenothiazine, a new commercially viable substituent for hole transport materials in perovskite solar cells. *J. Mater. Chem. C* **7**, 8593–8598 (2019).

25. Ullah, A. *et al.* Novel Phenothiazine-Based Self-Assembled Monolayer as a Hole Selective Contact for Highly Efficient and Stable p-i-n Perovskite Solar Cells. *Adv. Energy Mater.* **12**, 2103175 (2022).

26. Zhang, H. *et al.* Formamidinium Lead Iodide-Based Inverted Perovskite Solar Cells with Efficiency over 25 % Enabled by An Amphiphilic Molecular Hole-Transporter. *Angew. Chem. Int. Ed.* **63**, e202401260 (2024).

27. Planells, M., Abate, A., Snaith, H. J. & Robertson, N. Oligothiophene Interlayer Effect on Photocurrent Generation for Hybrid $TiO_2$ /P3HT Solar Cells. *ACS Appl. Mater. Interfaces* **6**, 17226–17235 (2014).

28. Liu, M. *et al.* Defect-Passivating and Stable Benzothiophene-Based Self-Assembled Monolayer for High-Performance Inverted Perovskite Solar Cells. *Adv. Energy Mater.* **14**, 2303742 (2024).

29. Chen, X. *et al.* Studies on the Effect of Solvents on Self-Assembled Monolayers Formed from Organophosphonic Acids on Indium Tin Oxide. *Langmuir* **28**, 9487–9495 (2012).

30. Paniagua, S. A., Li, E. L. & Marder, S. R. Adsorption studies of a phosphonic acid on ITO: film coverage, purity, and induced electronic structure changes. *Phys. Chem. Chem. Phys.* **16**, 2874 (2014).

31. Cui, D. *et al.* Making Room for Growing Oriented $FASnI_3$ with Large Grains via Cold Precursor Solution. *Adv. Funct. Mater.* **31**, 2100931 (2021).

32. Kaiser, M. *et al.* Interpreting the Time-Resolved Photoluminescence of Quasi-2D Perovskites. *Adv. Mater. Interfaces* **8**, 2101326 (2021).

33. Du, Y. *et al.* Manipulating the Formation of 2D/3D Heterostructure in Stable High-Performance Printable $CsPbI_3$ Perovskite Solar Cells. *Adv. Mater.* **35**, 2206451 (2023).

34. Zhang, M. *et al.* Ultralong photoluminescence lifetime enables efficient tin halide perovskite solar cells. *Mater.* **6**, 100425 (2025).

35. Liu, G. *et al.* Halogen Radical-Activated Perovskite-Substrate Buried Heterointerface for Achieving Hole Transport Layer-Free Tin-Based Solar Cells with Efficiencies Surpassing 14 %. *Angew. Chem.* **137**, e202419183 (2025).

36. Pascual, J. *et al.* Lights and Shadows of DMSO as Solvent for Tin Halide Perovskites. *Chem. – Eur. J.* **28**, e202103919 (2022).

37. Saidaminov, M. I. *et al.* Conventional Solvent Oxidizes Sn(II) in Perovskite Inks. *ACS Energy Lett.* **5**, 1153–1155 (2020).

38. Alias, N. *et al.* Photoelectrical Dynamics Uplift in Perovskite Solar Cells by Atoms Thick 2D $TiS_2$ Layer Passivation of $TiO_2$ Nanograss Electron Transport Layer. *ACS Appl. Mater. Interfaces* **13**, 3051–3061 (2021).

39. Caprioglio, P. *et al.* On the Origin of the Ideality Factor in Perovskite Solar Cells. *Adv. Energy Mater.* **10**, 2000502 (2020).

40. Aldamasy, M. H. *et al.* Photovoltaic potential of tin perovskites revealed through layer-by- layer investigation of optoelectronic and charge transport properties.

# Table of Contents

In this work, we develop Th-2EPT, a novel self-assembled monolayer (SAM), with optimized coordination strength and lattice compatibility for DMSO-free tin perovskite solar cells. Several advanced optical characterization techniques confirmed the higher quality of the Th-2EPT buried interface when compared to PEDOT and MeO-2PACz. Solar cells fabricated with Th-2EPT as the hole-selective layer achieved higher performance than PEDOT, the current state-of-the-art hole-selective material for tin perovskite.

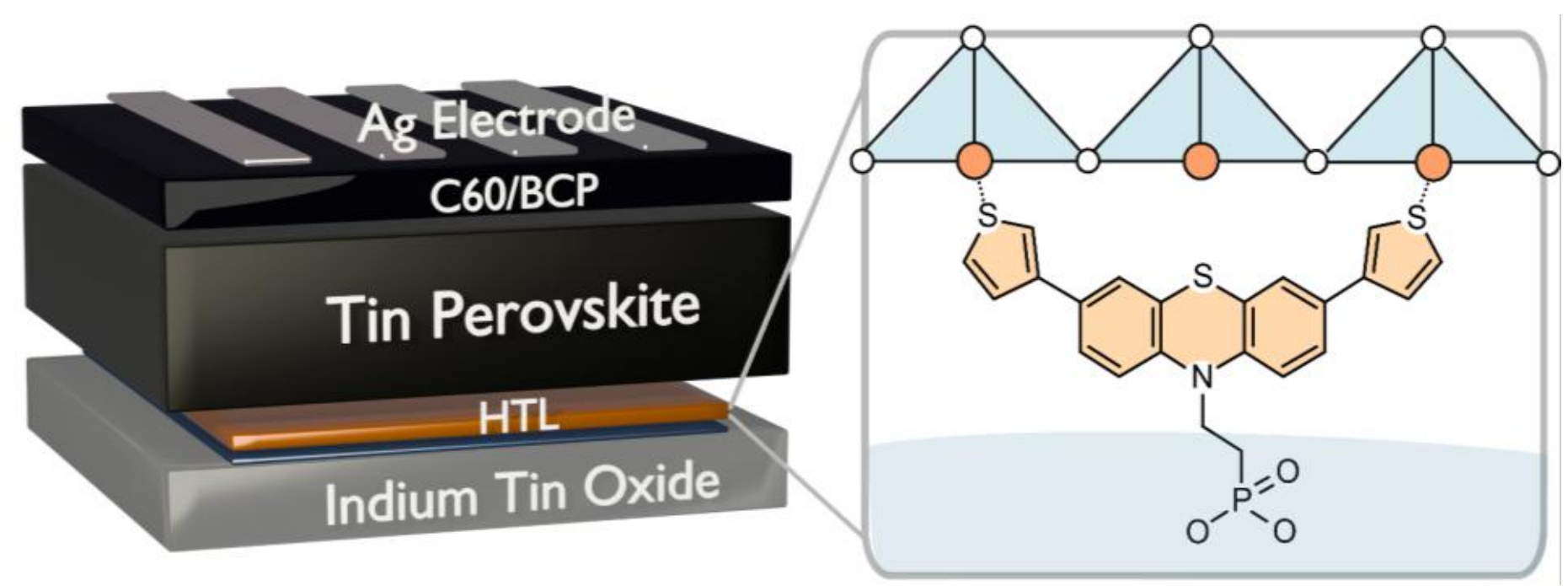

# Supporting Information
# For
# Phenothiazine-Based Self-Assembled Monolayer with Thiophene Head Groups Minimizes Buried Interface Losses in Tin Perovskite Solar Cells

**Authors**: Valerio Stacchini[§ 1,3], Madineh Rastgoo[§ 2], Mantas Marčinskas[5], Chiara Frasca[1,3], Kazuki Morita[1], Lennart Frohloff[4], Antonella Treglia[6], Thomas W. Gries, Orestis Karalis[1], Vytautas Getautis[5], Florian Ruske[1], Annamaria Petrozza[6], Norbert Koch[1,4], Hannes Hempel[1], Tadas Malinauskas[5*], Antonio Abate[1,2,3*], Artem Musiienko[7*]

**Affiliations**:

1. Helmholtz-Zentrum Berlin für Materialien und Energie GmbH, 12489 Berlin, Germany
2. Department of Chemical, Materials and Production Engineering, University of Naples Federico II, Fuorigrotta, Italy
3. Department of Chemistry, University of Bielefeld, Universitätsstraße 25, 33615 Bielefeld, Germany
4. Institut für Physik & Center for the Science of Materials (CSMB), Humboldt-Universität zu Berlin, 12489, Berlin

5. Department of Organic Chemistry, Kaunas University of Technology, Radvilenu pl. 19, Kaunas LT-50254, Lithuania
6. Istituto Italiano di Tecnologia (IIT), via Rubattino 81, 20134, Milan, Italy
7. Young Investigator Group, Robotized Material and Photovoltaic Engineering, Helmholtz-Zentrum Berlin für Materialien und Energie (HZB), Berlin, Germany.
§ These authors contributed equally to this work
* Corresponding authors: artem.musiienko@helmholtz-berlin.de, antonio.abate@helmholtz-berlin.de, tadas.malinauskas@ktu.lt

## Materials and Devices

# Chemicals

ITO substrates were purchased from Ossila. Water-free PEDOT complex dispersion in toluene (HTL3) was acquired from Clevios™. Formamidinium iodide was obtained from Dyenamo. Silver shots were supplied by Alfa Aesar and C60 was purchased from CreaPhys. All other chemicals, including $SnI_2$, $Al_2O_3$, Ethane-1,2-diammonium iodide, Bathocuproine (BCP), DEF (N,N-diethylformamide), DMPU (N,N′-dimethylpropyleneurea), N,N-dimethylformamide (DMF), toluene, ethanol, and diethyl ether (DEE), were provided by Sigma-Aldrich.

# Solar cell preparation

The ITO substrates underwent a sequential cleaning process through sonication, starting with a 2% Hellmanex solution in deionized water, followed by deionized water, acetone, and 2-propanol, each sonicated for 15 minutes. Then, the substrates were subjected to a UV-ozone treatment for 30 minutes and moved into a nitrogen-filled glovebox for further processing. For Th-2EPT dip coating deposition, following the UV-ozone treatment, the substrates were placed in a petri dish containing a SAM solution inside the glovebox (0.2 mM), and the lid was then closed. The substrates were left in the solution for 12 hours. After this period, the substrates were removed, spin-coated at 6000 rpm for 30 seconds, and annealed at 150 °C for 15 minutes. For the spin-coating deposition, a 2 mM stock solution was prepared in DMF. Th-2EPT solution (100 µL) was applied to the pre-cleaned ITO substrates, and after a 10-second delay, the substrates were spin-coated at 6000 rpm for 30 seconds. Subsequently, the SAM-coated substrates were annealed at 150 °C for 15 minutes. For the PEDOT devices, a diluted PEDOT-complex (100 µL, diluted 1:6 v in dry toluene) was dynamically deposited onto the ITO substrates and annealed at 150 °C for 10 minutes. To prepare the absorber layer, stock solutions of $SnI_2$ (1.2 M) and $EDAI_2$ (1 M) were prepared in a solvent system consisting of N,N-diethylformamide (DEF) and N,N'-dimethylpropyleneurea (DMPU) in a 1:6 volume ratio. These solutions were left overnight in a shaker at 20 °C. FAI powder was wiughted in a vial and mixed with the SnI2 stock solution in the appropriate amount to reach 1:1.1 FAI:SnI2 stoichiometry. Finally, the perovskite final solution was prepared by mixing $FASnI_3$ with 5% $EDAI_2$. This solution was then spin-coated onto the substrates, first at 500 rpm for 5 seconds and then at 4000 rpm for 40 seconds. At the 22-second, 100 µL of diethyl ether (DEE) was added to the spinning substrate to promote perovskite crystallization. The resulting film was annealed at 100 °C for 30 minutes. The device fabrication was completed by evaporating layers of C60 (40 nm), BCP (7 nm), and finally a layer of silver (140 nm).

# Methods

## Density Functional Theory (DFT) Calculations

The all-electron numeric-atom-centred orbital code FHI-aims was used to carry out the DFT calculations[1]. The molecular structure of MeO-2PACz and Th-2EPT was determined by performing DFT structure optimization with B3LYP exchange-correlation functional[2]. For $FASnI_3$, α-phase $FASnI_3$ was relaxed with reciprocal space sampling of 5 × 5 × 5. The molecular calculation and the bulk calculation were performed with open and periodic boundary conditions, respectively. The relaxed bulk $FASnI_3$ was then used to construct a $SnI_2$ terminated (001) surface slab model with an area of 4 × 3-unit cells and four layers in thickness. With periodic boundary conditions, the bottom two layers were fixed, and the rest was allowed to relax. From the observation that the anchoring group is bound to ITO and does not fold back as in the stable conformers, we replaced the anchoring group beyond phosphorus with hydrogen. The SAM on the surface were modelled by placing MeO-2PACz and Th-2EPT so that oxygen and sulfur were over the surface I-site, respectively. Since the distance between the chalcogens did not exactly match the I–I distance, we calculated two different bonding configurations each (Figure S1). During the relaxation, both SAM and surface were allowed to relax, and the calculation was performed with reciprocal space sampling of 1 × 1 × 1 and B3LYP functional. For the calculation of binding energy, we closely followed the definition: the energy required to remove the molecule from the surface.  We therefore calculated the energy difference between the SAM binded to the surface and when they are isolated.. To check the effect of dipole within the calculation cell, we performed the same calculation with dipole correction and confirmed that the effect of dipole is negligible. Furthermore, basis set superposition error (BSSE) for B3LYP is smaller than in post-Hartree-Fock methods (e.g. Møller–Plesset methods)[3], therefore BSSE correction was omitted.

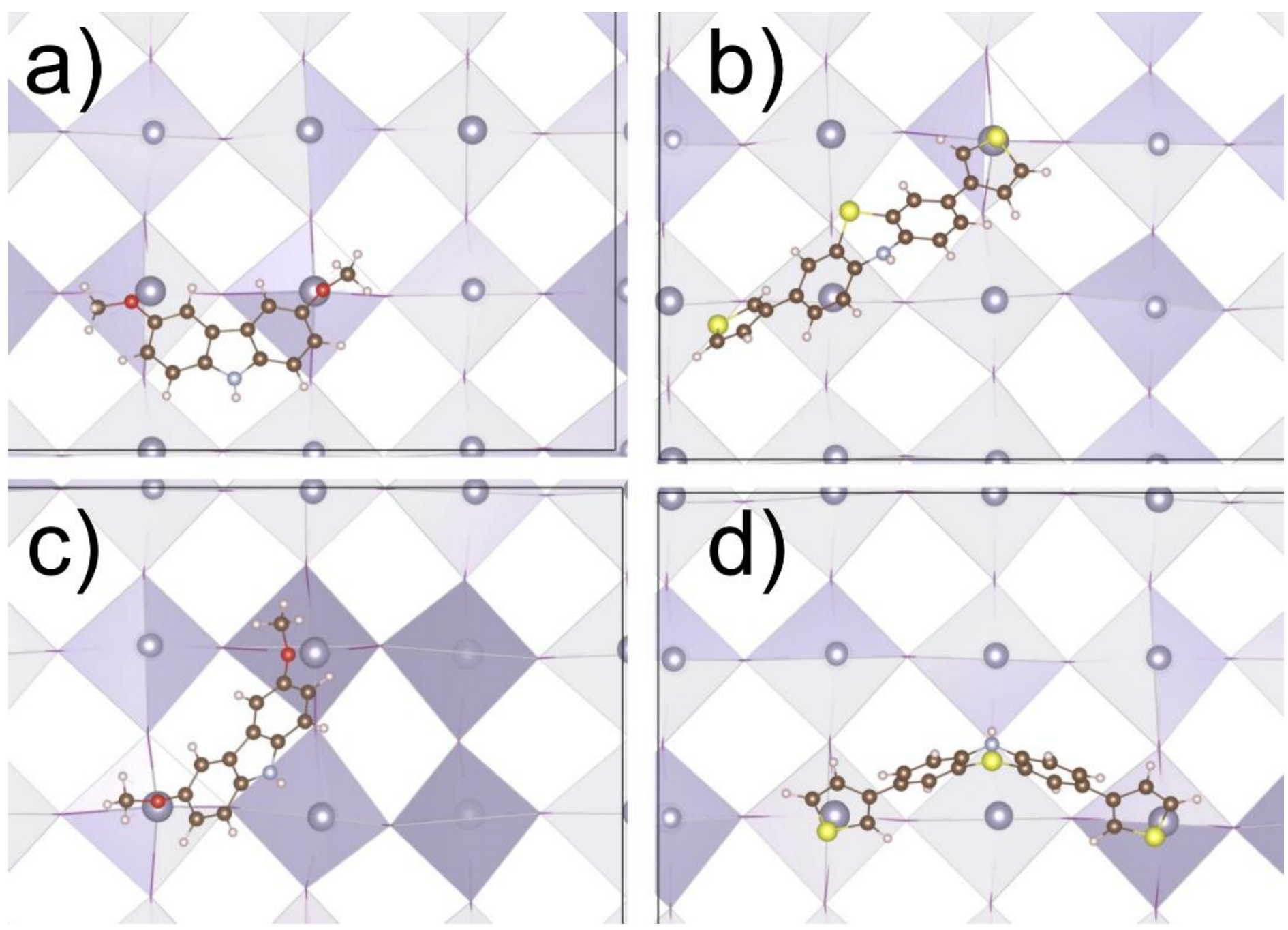

**Figure S1** DFT optimized molecular structure of MeO-2PACz **(a)** and **(c)**, and Th-2EPT **(b)** and **(d)** on $FASnI_3$ surface.

# Lattice Matching of SAMs

To assess the geometric compatibility between the SAM molecules and the $FASnI_3$ surface ($a \approx 6.3$ Å for $FASnI_3$), we evaluated the lateral distance between their passivating atoms (O–O in MeO-2PACz and S–S in Th-2EPT) and compared these with projected interatomic spacings of Sn atoms at the $SnI_2$-terminated (100) surface of cubic $FASnI_3$. The α-phase of $FASnI_3$ adopts a pseudo-cubic perovskite structure (space group $Pm\overline{3}m$), where tin atoms occupy the corner positions of a simple cubic lattice. Starting from a reference Sn atom at (0,0,0), the positions of its neighbors can be defined by integer lattice vector displacements in 3D space, and the interatomic distances can be calculated using the Euclidean norm. For example, the first-nearest neighbors lie at positions like (a,0,0), (0,a,0), (0,0,a), with a distance of *a*, the cubic lattice constant. Higher-order neighbors are located along face diagonals (e.g., $\sqrt{2}$a), body diagonals ($\sqrt{3}$a), and further out. We computed the Sn–Sn distances up to the fifth-nearest neighbors using this approach and projected these vectors onto the (100) surface to simulate the surface lattice viewed from above. The O–O distance in MeO-2PACz is approximately 7.81 Å, which best aligns with the second-nearest Sn–Sn spacing of ~9.22 Å, resulting in a geometric match of ~85%. In contrast, Th-2EPT has a head group S–S spacing of ~14.01 Å, which aligns closely with the fifth-nearest Sn–Sn distance (~14.6 Å), achieving a 96% match. The better geometric compatibility of Th-2EPT with the natural periodicity of the surface suggests reduced interfacial strain and improved lattice templating. A full table of calculated Sn–Sn spacings, neighbor indices, and match percentages is provided in Figure S2, along with a visual representation of neighbor shells in the 3D perovskite lattice.

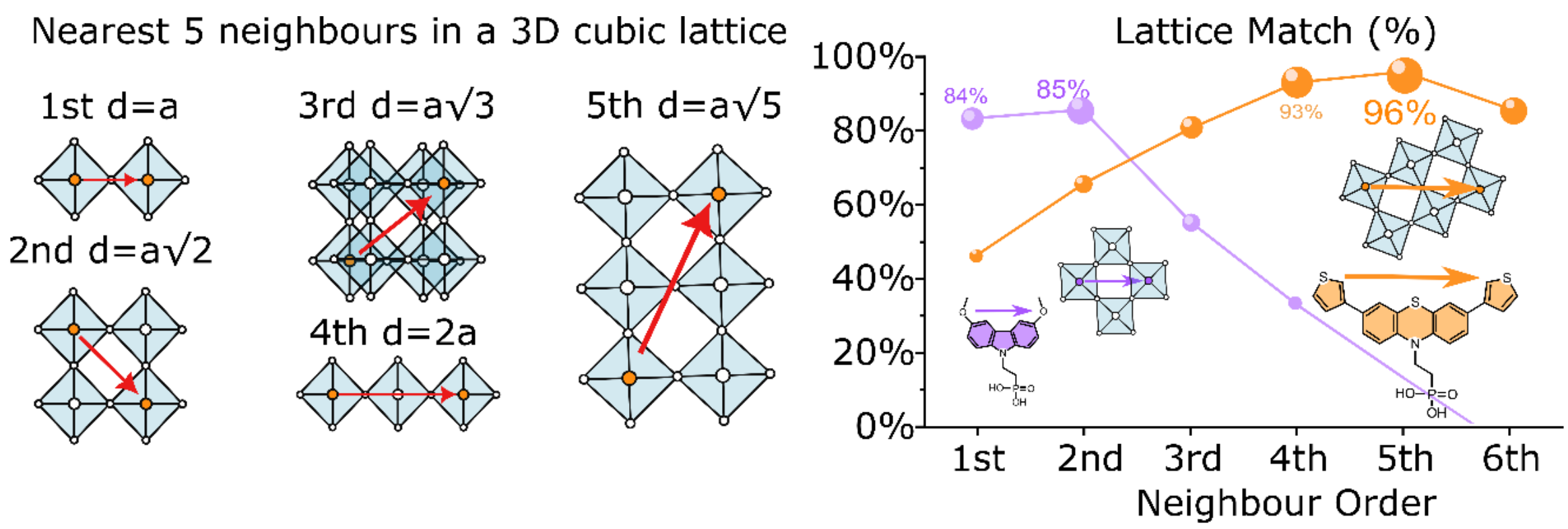

| Neighbor Order | Multiple of a | Distance (Å) | Match with MeO-2PACz | Match with Th-2EPT |
|---|---|---|---|---|
| 1st | $\sqrt{(1^2+0^2+0^2)}$ | 6,52 | 83.5% | 46.5% |
| 2nd | $\sqrt{(1^2+1^2+0^2)}$ | 9,22 | **84.7%** | 65.8% |
| 3rd | $\sqrt{(1^2+1^2+1^2)}$ | 11,29 | 69.2% | 80.6% |
| 4th | $\sqrt{(2^2+0^2+0^2)}$ | 13,04 | 59.9% | 93.1% |
| 5th | $\sqrt{(2^2+1^2+0^2)}$ | 14,58 | 53.6% | **96.1%** |

**Figure S2** Lattice match calculations. In the upper left figure, a graphical representation of lattice vectors corresponding to the first 5 nearest neighbouring Sn ions in the $FASnI_3$ cubic tridiemnsional lattice. On the right, a plot represents Lattice match as a function of the neighbouring order. Below, the same data is reported in a table.

# X-Ray Diffraction (XRD)

XRD was carried out on a Bruker D8 diffractometer in Bragg–Brentano geometry, using Cu Kα radiation (λ = 1.5406 Å), 40 kV acceleration voltage, and 40 mA current. Samples were measured under inert conditions using airtight poly(methyl methacrylate) (PMMA) sample holders by Bruker.

To remove background noise and any sloping baseline, an asymmetric least squares (ALS) algorithm was applied to each dataset. To aid in peak visual clarity and detection, a mild smoothing filter was applied to the baseline-corrected data. A simple moving average (window ~11 points, corresponding to ~0.22°) was used to reduce high-frequency noise. All peak-finding operations were performed on these smoothed intensity curves. (Notably, smoothing was only used for identifying and plotting peaks; the quantitative peak fitting for obtaining FWHM employed the original unsmoothed data to preserve accuracy). An automated peak-finding algorithm (SciPy's find_peaks) scanned each smoothed XRD pattern to identify significant diffraction peaks. A Gaussian fit of the (100) peak for each sample yielded the peak position (around 14.3° 2θ) and the **FWHM** (full width at half maximum) in degrees. The table below summarizes the measured peak position and FWHM for the (100) reflection of each sample. All three samples have essentially the same peak position within ~0.1°, but their FWHM values differ significantly.

| **Sample** | **(100) Peak Position *(° 2θ)*** | **FWHM *(° 2θ)*** |
| --- | --- | --- |
| **PEDOT** | 14.38° | 0.10° |
| **MeO** | 14.30° | 0.30° |
| **Th-2EPT** | 14.30° | 0.12° |

# Scanning Electron Microscope (SEM) Imaging

SEM images were collected using a Zeiss Merlin SEM with a Gemini 2 column. Samples were transferred from the glovebox to the SEM inside an airtight transfer shuttle in inert ambient. Images from the top (figure 2) were collected with an acceleration voltage of 2 kV and recorded with an Everhart-Thornley detector to highlight the topological contrast.

# UV-Vis Absorbance Spectra and Photoluminescence Quantum Yield (PLQY)

UV-vis steady state spectra were measured on perovskite thin films deposited on bare glass using a UV/VIS/NIR spectrophotometer Lambda 1050, PerkinElmer, in the wavelength range 350–1100 nm, a step size of 1 nm. The excitation density dependent PLQY measurement was performed by illuminating the sample from the glass side with a femtosecond laser (Light conversion Pharos) at 500 kHz and wavelength of 343 nm and with variable power density. The photoluminescence was collected in reflection geometry with a Maya 2000 Pro visible spectrometer.

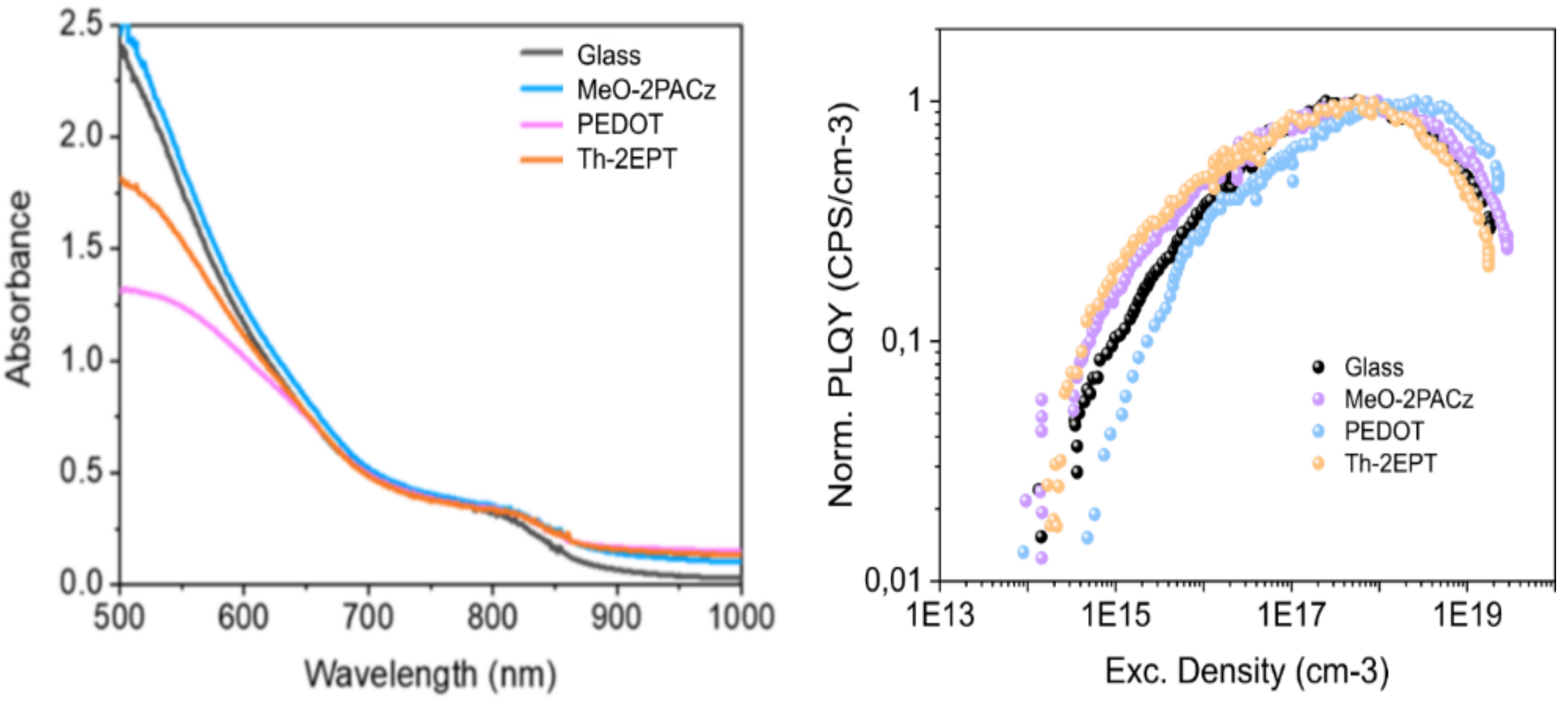

**Figure S3** Absorbance spectra (UV-vis) of thin films of tin perovskite coated on top the different HTL substrates. Normalized PLQY obtained from the same samples.

# Transient Absorption Spectroscopy (TA)

For TAS characterization an amplified femtosecond laser (Light Conversion Pharos) generated pulses of ~280 fs centred at 1030 nm with a repetition rate of 2 kHz. A broadband white light probe is generated by focusing the pulses into a thin sapphire plate. At short delays (<5 ns), the third harmonic of the fundamental provided the pump light (343 nm). At long delays (>1 ns), pump light at 354 nm was provided by the second harmonic of a Q-switched Nd:Yag laser (Innolas Picolo), which was electronically triggered and synchronized to the femtosecond laser via an electronic delay. The pump excitation density is $4 \cdot 10^{17}$ $cm^{-3}$.

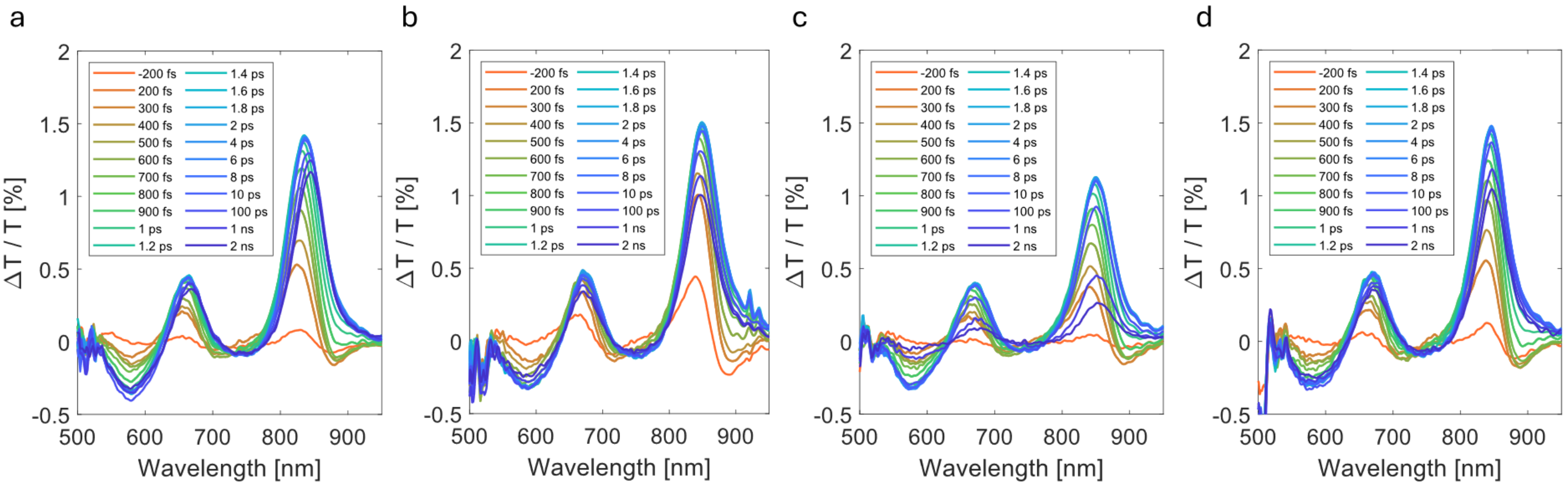

**Figure S4** Transient Absorption Spectra of perovskite on a) glass, b) MeO-2PACz, c) PEDOT, d) Th-2EPT as function of pump-probe delay

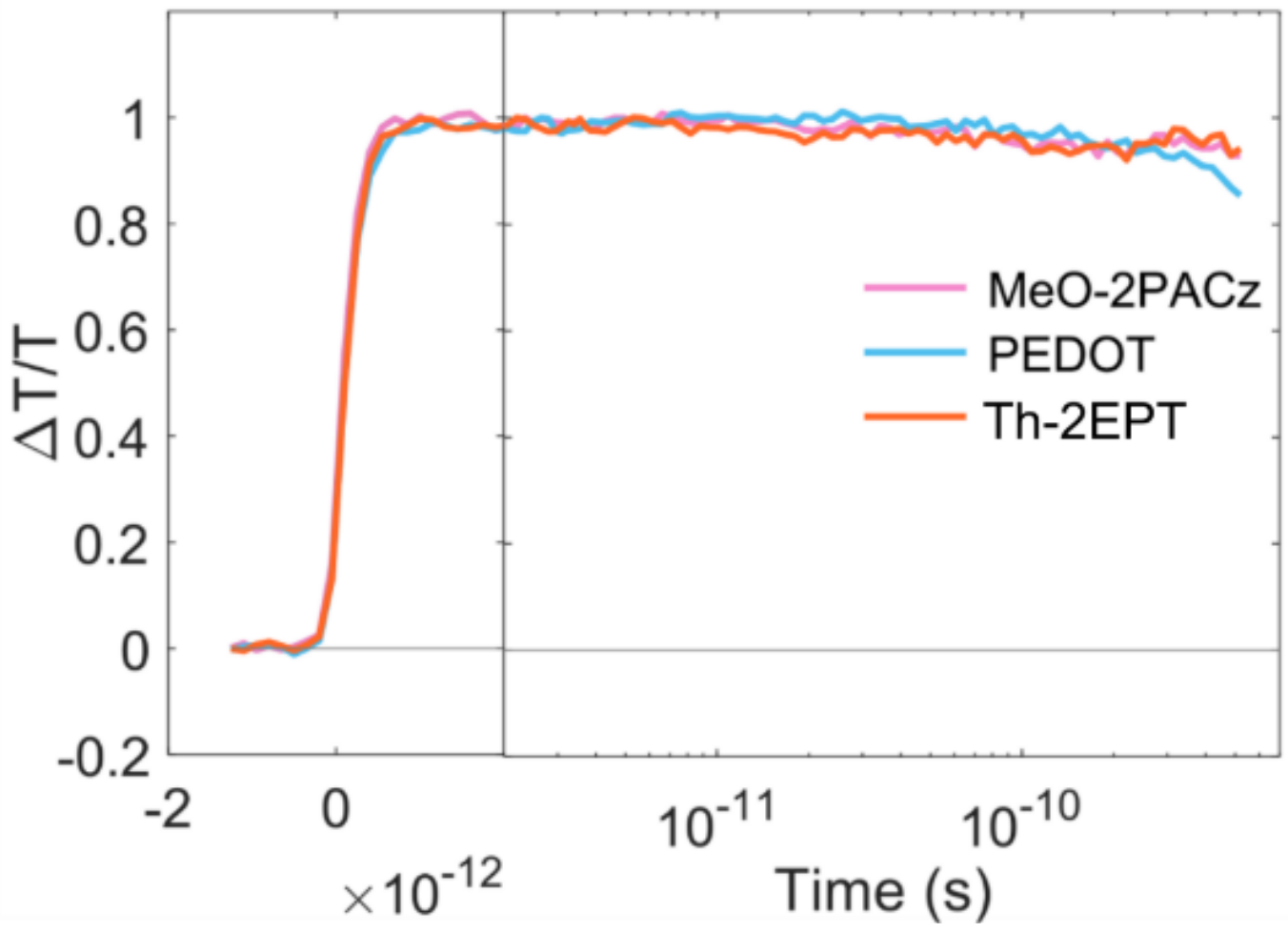

**Figure S5** Transient Absorption Spectra of perovskite on **a)** glass, **b)** MeO-2PACz, **c)** PEDOT, **d)** Th-2EPT as function of pump-probe delay

# Transient Photoluminescence (trPL)

Tr-PL is a common technique for measuring charge carrier recombination in semiconductors. The samples were excited with a femtosecond laser of 515 nm wavelength and repetition rate of 125 kHz. The laser power was attenuated with neutral density filters to 2.4 pJ*cm$^{-2}$/pulse. This corresponds to a photoexcitation of $6 \cdot 10^{13}$ carriers*cm$^{-3}$. This carrier concentration is typical for such perovskite thin films to open circuit conditions under 1 sun illumination. The PL emission was measured by time-correlated single photon counting with PicoHarp300. To this end, a Geiger-mode avalanche photodiode was used with a 530 nm long-pass filter in front of it to suppress the 515 nm laser light.

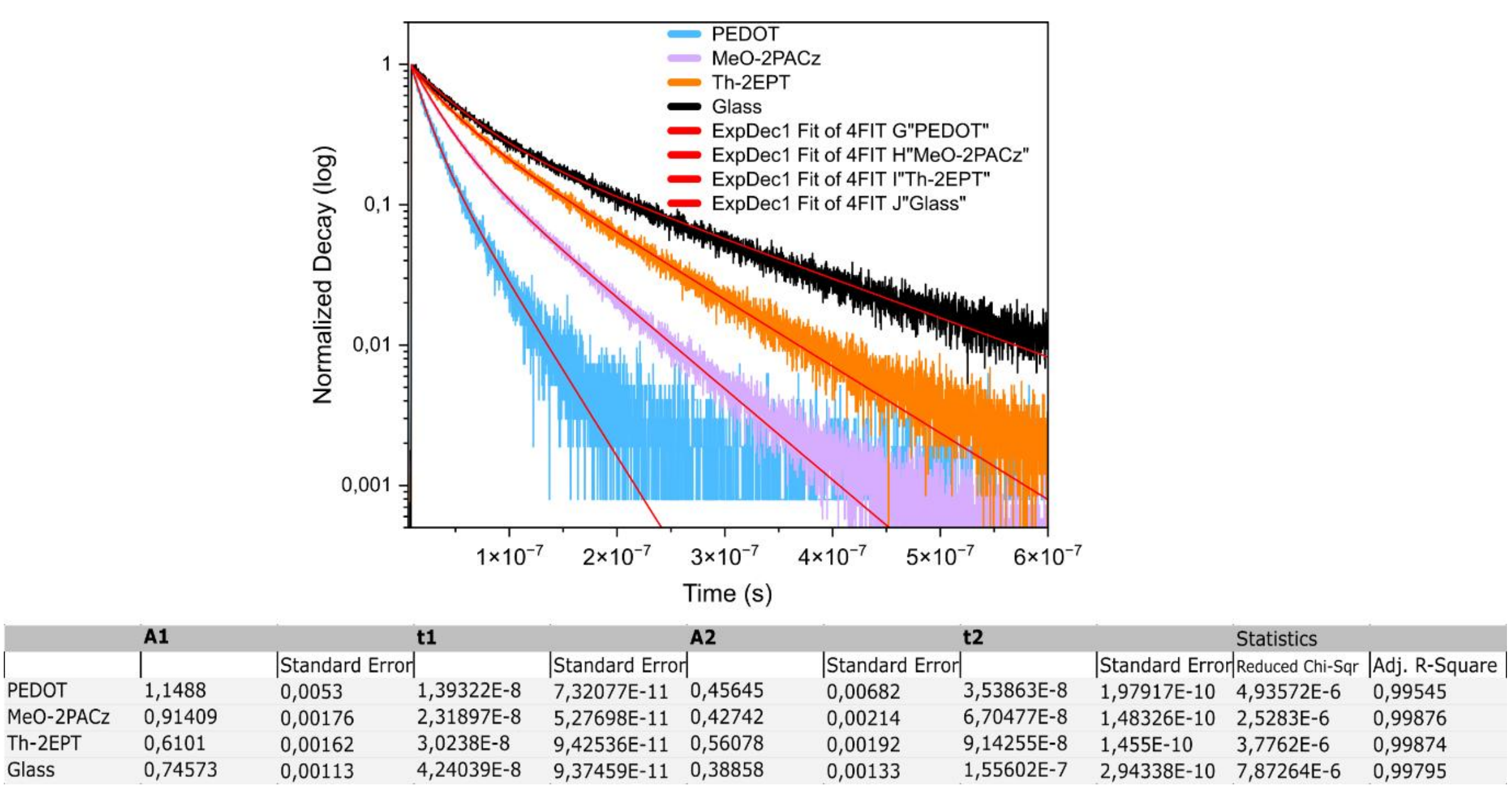

| | A1 | | t1 | | A2 | | t2 | | Statistics | |
|---|---|---|---|---|---|---|---|---|---|---|
| | | Standard Error | | Standard Error | | Standard Error | | Standard Error | Reduced Chi-Sqr | Adj. R-Square |
| PEDOT | 1,1488 | 0,0053 | 1,39322E-8 | 7,32077E-11 | 0,45645 | 0,00682 | 3,53863E-8 | 1,97917E-10 | 4,93572E-6 | 0,99545 |
| MeO-2PACz | 0,91409 | 0,00176 | 2,31897E-8 | 5,27698E-11 | 0,42742 | 0,00214 | 6,70477E-8 | 1,48326E-10 | 2,5283E-6 | 0,99876 |
| Th-2EPT | 0,6101 | 0,00162 | 3,0238E-8 | 9,42536E-11 | 0,56078 | 0,00192 | 9,14255E-8 | 1,455E-10 | 3,7762E-6 | 0,99874 |
| Glass | 0,74573 | 0,00113 | 4,24039E-8 | 9,37459E-11 | 0,38858 | 0,00133 | 1,55602E-7 | 2,94338E-10 | 7,87264E-6 | 0,99795 |

**Figure S6** Double exponential fitting parameters derived from the time-resolved photoluminescence (trPL) measurements. The plots display the raw trPL data with the corresponding fitted curves overlaid for comparison.

# Solar Cells

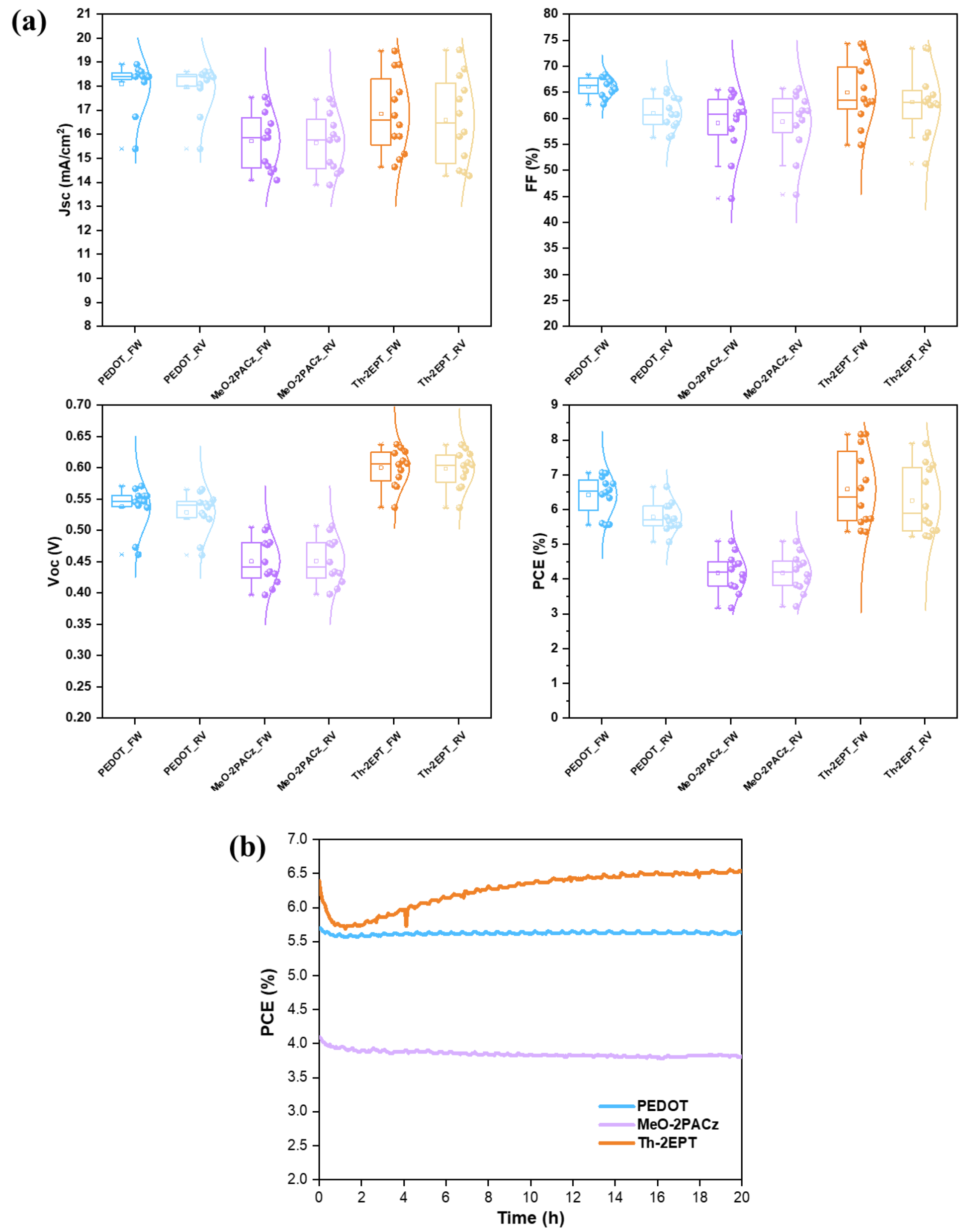

**Figure S7** (a) Jsc, FF, Voc, and PCE statistics from both forward and reverse JV scans of tin perovskite solar cells (TPSCs). (b) Short-term operational stability of TPSC devices under continuous one-sun illumination in a nitrogen atmosphere.

We compared the operational stability of devices with different HSLs under maximum power point (MPP) tracking in a nitrogen atmosphere. The devices based on PEDOT and MeO-2PACz showed similarly stable performance over 20 hours. In contrast, the Th-2EPT device initially exhibited a decline in performance during the first 2 hours, followed by a noticeable recovery as the test progressed.

# Contact Angle Measurement

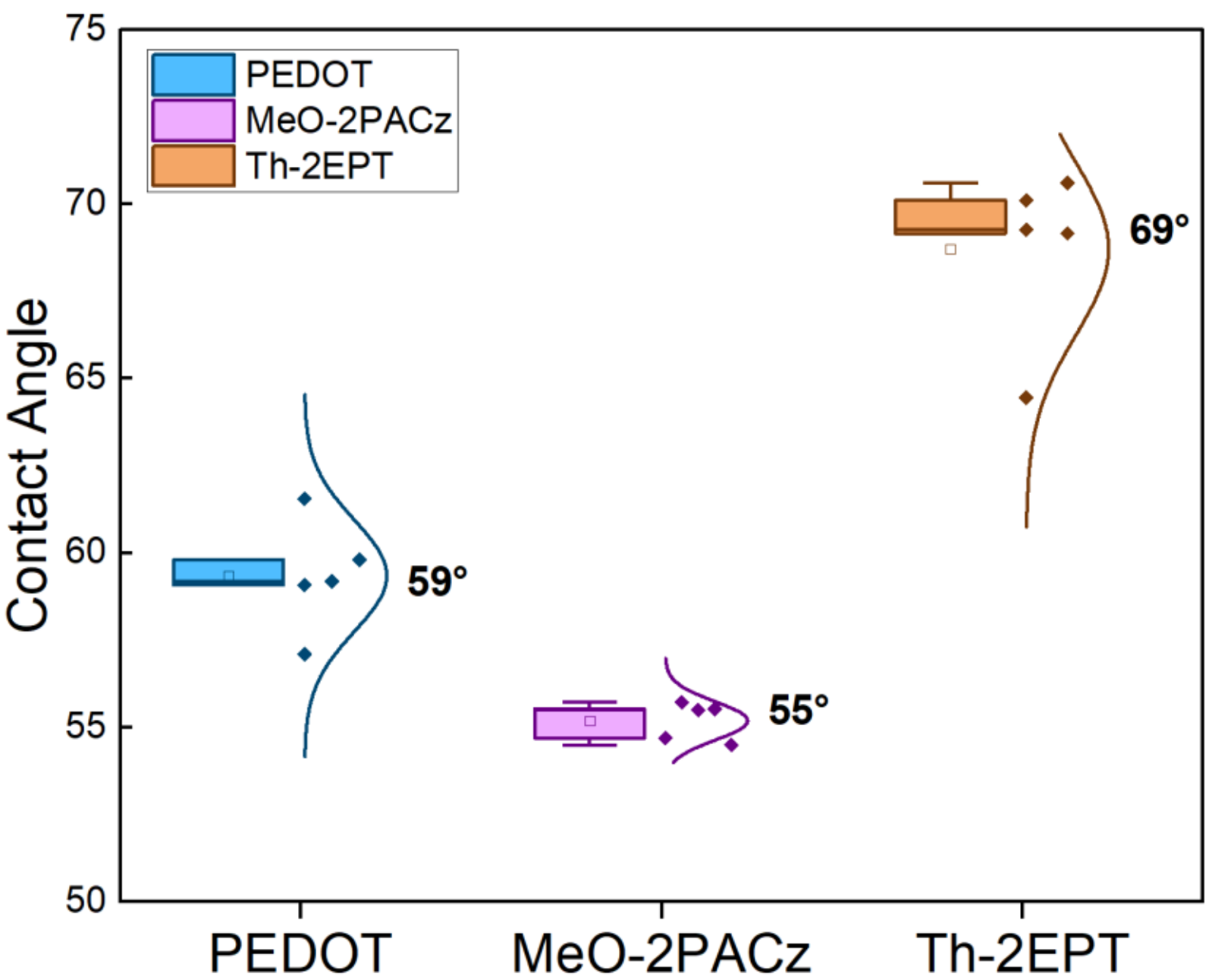

**Figure S8** Contact angle measurements for PEDOT, MeO-2PACz, and Th-2EPT layers deposited on ITO substrates. Th-2EPT exhibits a contact angle of 69°. PEDOT achieves a contact angle of 59°, aided by an alumina nanoparticle interlayer. MeO-2PACz displays the lowest contact angle (55°).

# Ultraviolet Photoemission Spectra (UPS) and X-ray Photoemission Spectra (XPS)

The ultraviolet photoemission spectra were recorded using a hemispherical electron analyser (SPECS Phoibos 100) in a UHV system equipped with a monochromated He I (hν = 21.218 eV) source. The selected pass energy 5 eV offers a setup resolution of 105 meV as determined from the Fermi edge of sputter-annealed gold. The base pressure of the analysis chamber was 3 x 10-10 mbar. All spectra were recorded under normal emission with an acceptance angle of ca. 6°. For measurement of the SECO, a sample bias of -10.0 V was employed to overcome the analyser work function. The low-energy inverse photoemission spectra were recorded using a LEIPES system (ADCAP Ltd.) consisting of an electron gun with a BaO cathode and a Hitachi photomultiplier tube behind a 250±6 nm bandpass filter in a UHV chamber with base pressure better than 1 x 10-9 mbar. The setup resolution was determined to be 600 meV. The x-ray photoemission spectra were acquired with a JEOL JPS-9030 with a vacuum base pressure of 3 x 10-9 mbar using monochromated Al Kα radiation (hν = 1486.6 eV) for excitation. The spectra were recorded with pass energy 20 eV, resulting in a setup resolution of 820 meV as determined from the FWHM of the Ag 3d5/2 peak and under an emission angle of ~10°. All spectra were recorded at room temperature. In order to establish the energy level alignment at the FASI/Th-2EPT and FASI/PEDOT interfaces, photoemission spectroscopy measurements were performed. The samples were prepared ex-situ in a nitrogen-filled glovebox and transferred into the UHV system without air exposure. All measurements were performed within three days after sample fabrication.

# UPS characterisation

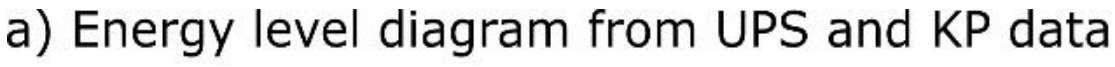

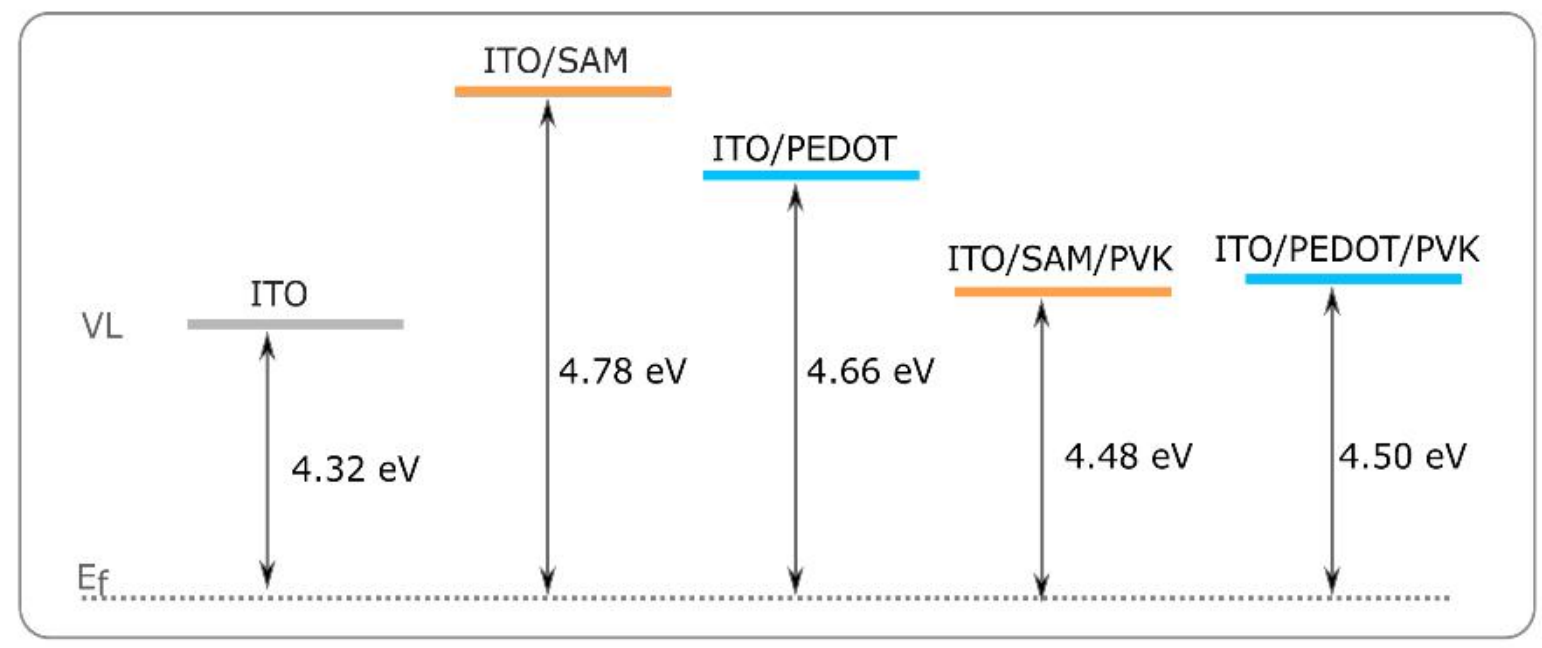

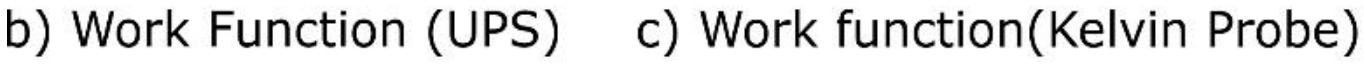

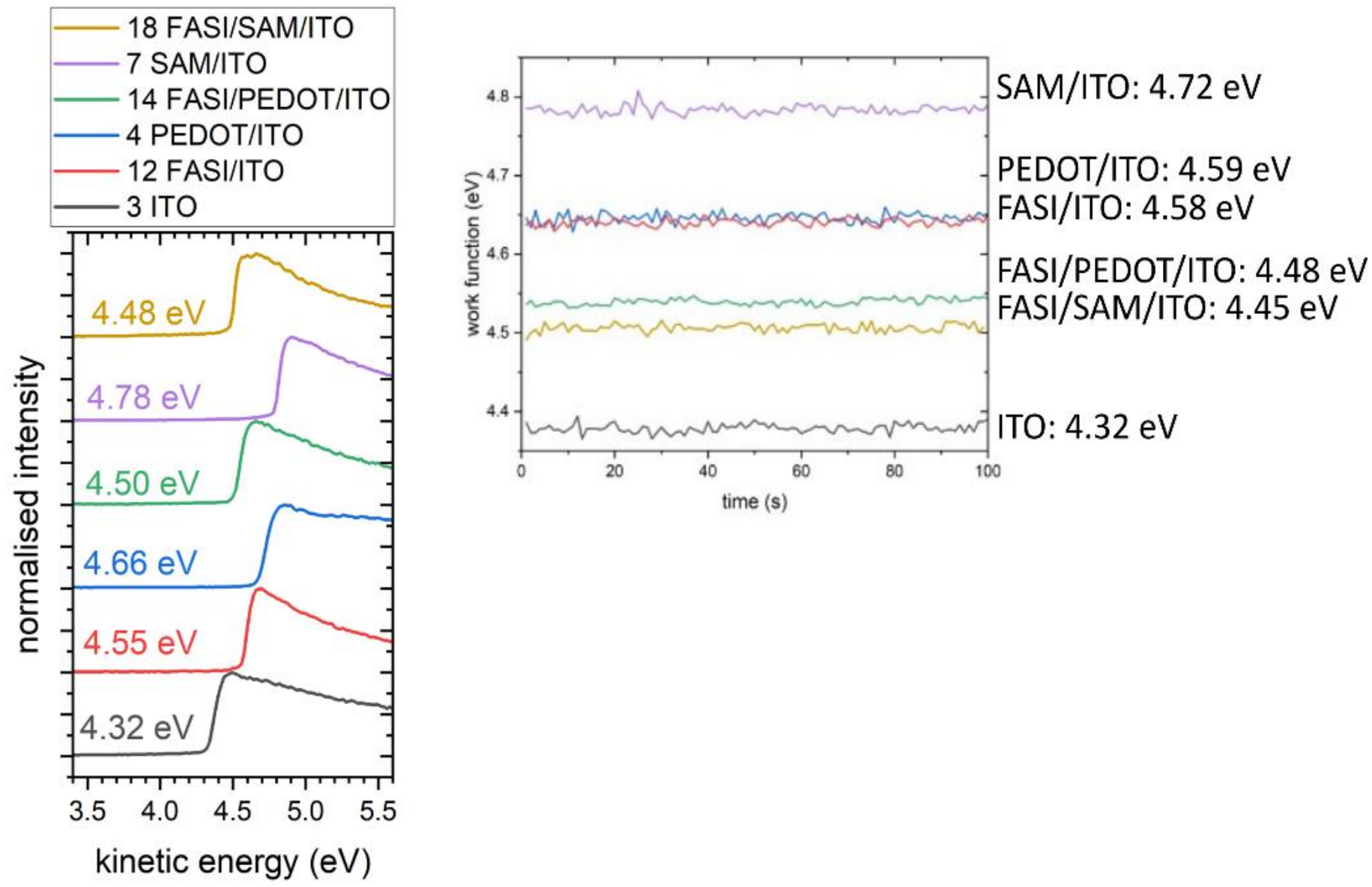

**Figure S9** energy level diagram **(a)** obtained from UPS Measurements for kinetic and binding energy **(b)** and Kelvin Probe for work function values **(c)**.

In Figure S9 the secondary electron cut-offs of the ITO substrate, HTL/ITO and FASI/HTL/ITO are shown for PEDOT and Th-2EPT SAM. The work function of the ITO substrate was determined to be 4.32 eV and slightly increases to 4.78 eV and 4.66 eV after deposition of the SAM or PEDOT, respectively. The increase of the work function upon deposition of the SAM can be rationalised by means of the dipole character of the Th-2EPT SAM[4]. The electron attracting nature of the thiophene groups causes a molecular dipole pointing away from the sample surface and thereby provides an additional barrier for electron emission, consequently slightly increasing the work function. This effect can be quantified by the means of the Helmholtz equation:

$$\Delta\Phi = \frac{e\mu_{\perp}}{\epsilon_0 \epsilon_r}\frac{N}{A} = \frac{e\mu \cos\theta}{\epsilon_0 \epsilon_r}\frac{N}{A}$$

The stronger the dipole moment of the molecule and the higher the molecule density per area, the higher the work function change upon SAM deposition. [Inserting $\Delta\Phi$ = 0.46 eV, assuming a dipole moment of $\mu$ = 0.61 eÅ as calculated from DFT and a dielectric constant $\epsilon_r$ = 2,[4] that gives the tilt angle of the molecule $\theta$ and the molecule density per area $N/A$ as unknown parameters. For hypothetical tilt angle of 0° (perpendicular to surface) this gives N/A = 0.83 $nm^{-2}$. Calculating for other angles: theta = 30° gives N/A = 0.96 $nm^{-2}$, theta = 45° gives N/A = 1.18 $nm^{-2}$. For comparison: 1 $nm^{-2}$ is approximately one molecule per ITO unit cell[5]. After deposition of FASI, all samples show a work function around 4.50 eV. The independence of the FASI energy levels concerning the substrate work function points to pinning of the Fermi level. The strong p-type character matches up with the well-known self-doping in tin perovskites.[6] The slight positive barrier for hole extraction should, however, not limit the functionality of the solar cell as the SAM layer is only less than one nanometre thin. All work functions measured through UPS were further confirmed with vacuum Kelvin Probe measurements to exclude any influence by the high-energy radiation required for UPS. Further, no surface photo voltage greater than 50 meV was detected for any sample at additional 1.5 suns white light illumination during UPS measurements. In Fig. S9, a schematic representation of the energy level alignments in the respective half-cells is depicted. For FASI, a negligible exciton binding energy is assumed as rationalised for other perovskites.

a) molecular structure of Th-2EPT

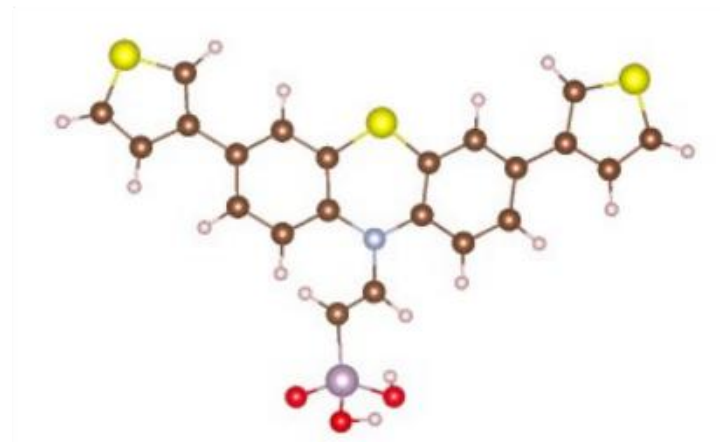

b) peak areas for different XPS signals

| Orbital | Area | RSF | Area/RSF |
|---|---|---|---|
| S 2p | 62 | 1.54 | 40.3 |
| P 2p | 16 | 1.11 | 14.4 |
| N 1s | 24 | 1.73 | 13.9 |
| C 1s | 328 | 1.00 | 328 |
| In 3d | 5473 | 10.90 | 502.1 |

c) Phosphorous, sulfur, nitrogen  specific peaks

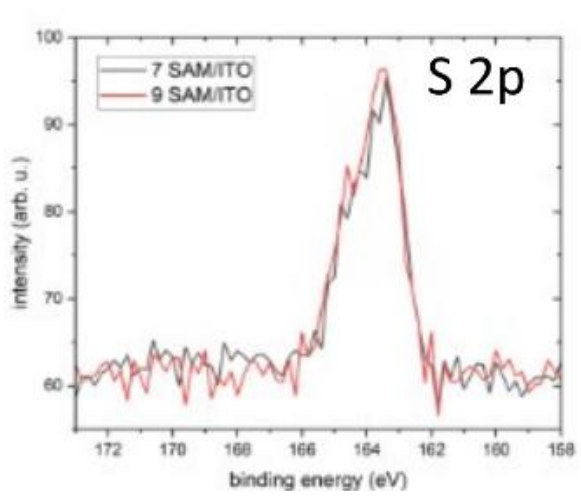

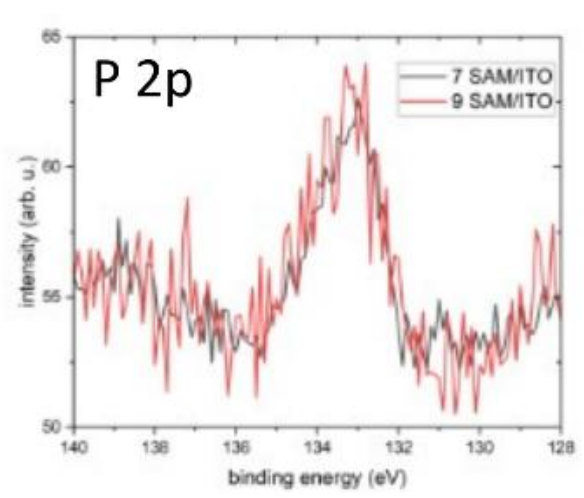

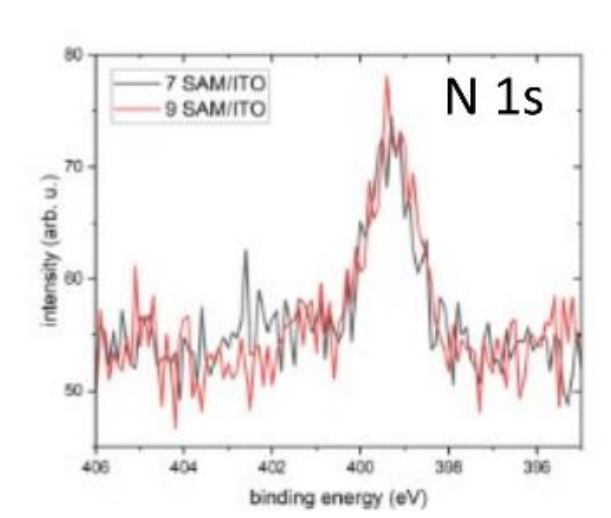

**Figure S10** XPS Spectra showing a model of the molecular structure from the software Avogadro; **b)** peak areas ratios for different elemental orbitals [RSF = Relative Sensitivity Factor]; **c)** specific peaks.

# XPS characterisation

The x-ray photoemission spectroscopy confirms the presence of the Th-2EPT molecule on the ITO surface. A quantitative analysis of the SAM/ITO data confirms the expected stoichiometry between P:S:N:C within the molecule to be 1.0:2.8:1.0:23 (expected: 1:3:1:22). The denticity of SAM binding could not be extracted from the measurement, but the present of the molecule on ITO is confirmed.

# Synthesis of Th-2EPT (V1495)

Chemicals were purchased from Sigma-Aldrich, TCI Europe, and used as received without further purification. $^{1}$H NMR spectra were recorded at 400 MHz on a Bruker Avance III spectrometer, $^{13}$C NMR spectra were collected using the same instrument at 101 MHz. The chemical shifts, expressed in ppm, are reported relative to tetramethylsilane (TMS). Reactions were monitored by thin-layer chromatography on ALUGRAM SIL G/UV254 plates and developed with UV light. Silica gel (grade 9385, 230–400 mesh, 60 Å, Aldrich) was used for column chromatography. Elemental analysis was performed with an Exeter Analytical CE-440 elemental analyzer, Model 440 C/H/N/. Electrothermal A.KRÜSS M3000 capillary melting point apparatus was used for determination of melting points.

### *Synthesis scheme*

10-*H*-Phenothiazine was alkylated using 1,2-dibromoethane. Alkylated derivative **1** was brominated using *N*-bromosuccinimide in DMF. Isolated phenothiazine **2** was used for the palladium catalyzed Suzuki-Miyaura coupling reaction with 3-thienylboronic acid under argon in anhydrous 1,4-dioxane, resulting in intermediate **3**, which was further transformed into phosphonate **4** by refluxing in triethyl phopshite. Final product Th-2EPT SAM molecule, containing phosphonic acid functional group, was obtained by hydrolysis of obtained intermediate, using bromotrimethylsilane, methanol and water.

NBS
DMF, 25 °C

KOH (50 %), TBAB,
80 °C

**1** (66 %)

**2** (81 %)

$Pd(PPh_3)_4$, 2M $K_2CO_3$,
1,4-dioxane, 70 °C

**3** (82 %)

reflux

**4** (78 %)

1. 1,4-dioxane,
25 °C
2. MeOH
3. $H_2O$

**V1495** (70 %)

**Scheme S1.** Synthesis route to the phenothiazine-based Th-2EPT SAM (or V1495) containing thiophenes and phosphonic acid anchoring group.

*Experimental procedure*

*10-(2-bromoethyl)-10H-phenothiazine (**1**)*

10*H*-Phenothiazine (5 g, 25.1 mmol) was dissolved in 1,2-dibromoethane (54 ml, 627 mmol), followed by addition of 50 % KOH aqueous solution (14 ml, 125 mmol) and tetrabutylammonium bromide (1.21 g, 3.76 mmol). Reaction mixture was heated to 80 °C and conducted for 96 hours. After first 24 hours additional 50 % KOH aqueous solution (14 ml, 125 mmol) and tetrabutylammonium bromide (1.21 g, 3.76 mmol) were added. After termination of reaction (TLC, eluent acetone:*n*-hexane, 3:22), organic components were extracted with ethyl acetate, organic layer dried over anhydrous $Na_2SO_4$, filtered and solvent removed under reduced pressure. Crude product was purified by column chromatography (eluent acetone:*n*-hexane, 3:22). Product obtained as yellow crystals (5.12 g, 66 % yield). M.p. 78–80 °C. $^{1}$H NMR (400 MHz, DMSO-$d_6$): δ 7.24 – 7.13 (m, 4H), 7.04 (d, *J* = 8.1 Hz, 2H), 6.97 (t, *J* = 7.4 Hz, 2H), 4.30 (t, *J* = 6.3 Hz, 2H), 3.74 (t, *J* = 6.3 Hz, 2H) ppm. $^{13}$C NMR (101 MHz, DMSO-d6): δ 143.96, 127.71, 127.26, 123.99, 122.91, 115.76, 48.36, 29.78 ppm. Anal. calcd. for $C_{14}H_{12}BrNS$: C 54.91, H, 3.95, N, 4.57; found: C 55.07, H 3.85, N 4.42.

*3,7-Dibromo-10-(2-bromoethyl)-10H-phenothiazine (**2**)*

10-(2-Bromoethyl)-10*H*-phenothiazine (**1**, 2 g, 6.53 mmol) was dissolved in DMF (25 ml) and afterwards *N*-bromosuccinimide (2.38 g, 13.4 mmol) was added portionwise. Reaction conducted overnight at 25 °C. After termination of reaction (TLC, eluent acetone:*n*-hexane, 1:24) organic components extracted with ethyl acetate, organic layer dried over anhydrous $Na_2SO_4$, filtered and solvent evaporated under reduced pressure. Crude product was purified by column chromatography (eluent acetone:*n*-hexane, 1:24), resulting in white crystals (2.44 g, 81 % yield) as a product. M.p. 144–145 °C.

$^{1}$H NMR (400 MHz, DMSO-$d_6$): δ 7.39 – 7.32 (m, 4H), 6.98 (d, *J* = 8.6 Hz, 2H), 4.26 (t, *J* = 6.2 Hz, 2H), 3.70 (t, *J* = 6.1 Hz, 2H) ppm. $^{13}$C NMR (101 MHz, DMSO-$d_6$): δ 143.05, 130.43, 129.21, 125.96, 117.70, 114.52, 48.49, 29.58 ppm. Anal. calcd. for $C_{14}H_{10}Br_3NS$: C 36.24, H, 2.17, N, 3.02; found: C 36.50, H 2.02, N 2.88.

*10-(2-Bromoethyl)-3,7-di(thiophen-3-yl)-10H-phenothiazine (**3**)*

3,7-Dibromo-10-(2-bromoethyl)-10*H*-phenothiazine (**2**, 1.5 g, 3.23 mmol) was dissolved in anhydrous 1,4-dioxane (40 ml) under argon atmosphere, followed by addition of 3-thienylboronic acid (1.03 g, 8.08 mmol), $Pd(PPh_3)_4$ (0.37 g, 0.32 mmol) and $K_2CO_3$ 2M aqueous solution (5.9 ml, 9.69 mmol). Reaction conducted at 80 °C under inert argon atmosphere for 24 hours. After termination of reaction (TLC, eluent acetone:*n*-hexane, 2:23), reaction mixture was cooled down and filtered through celite which was washed with THF. Organic solvent was evaporated, and the crude product was purified by column chromatography using acetone:*n*-hexane (2:23) as an eluent resulting yellow crystals (1.24 g, 82 %) as a product. M.p. 177.5–179 °C (melting and decomposition). $^{1}$H NMR (400 MHz, DMSO-$d_6$): δ 7.82 (d, *J* = 1.1 Hz, 2H), 7.62 – 7.58 (m, 2H), 7.57 – 7.48 (m, 6H), 7.10 – 7.01 (m, 2H), 4.35 (t, *J* = 6.1 Hz, 2H), 3.81 – 3.73 (m, 2H) ppm. $^{13}$C NMR (101 MHz, DMSO-$d_6$): δ 142.55, 140.20, 130.20, 127.03, 126.00, 125.44, 124.65, 124.06, 120.12, 115.95, 48.46, 29.86 ppm. Anal. calcd. for $C_{22}H_{16}BrNS_3$: C 56.17, H, 3.43, N, 2.98; found: C 56.40, H 3.30, N 2.71.

*Diethyl {2-[3,7-di(thiophen-3-yl)-10H-phenothiazin-10-yl]ethyl}phosphonate (**4**)*

Diethyl {2-[3,7-di(thiophen-3-yl)-10*H*-phenothiazin-10-yl]ethyl}phosphonate (**3**, 1.2 g, 2.55 mmol) was suspended in triethyl phosphite (3 ml, 51 mmol) and the reaction was refluxed overnight. After termination of reaction (TLC, eluent acetone:*n*-hexane, 7:18), solvent was removed under reduced pressure and the crude product was purified by column chromatography (eluent acetone:*n*-hexane, 7:18). Product obtained as yellow resin (1.05 g, 78 %). $^{1}$H NMR (400 MHz, DMSO-$d_6$): δ 7.83 (s, 2H), 7.63 – 7.52 (m, 8H), 7.05 (d, *J* = 8.4 Hz, 2H), 4.15 – 3.94 (m, 6H), 2.35 – 2.20 (m, 2H), 1.24 (t, *J* = 7.1 Hz, 6H) ppm. $^{13}$C NMR (101 MHz, DMSO-$d_6$): δ 142.43, 140.16, 130.00, 126.99, 125.96, 125.41, 124.51, 123.37, 120.04, 115.49, 61.35, 61.29, 40.93, 23.79, 22.45, 16.31, 16.26 ppm. Anal. calcd. for $C_{26}H_{26}NO_3PS_3$: C 59.18, H, 4.97, N, 2.65; found: C 59.41, H 4.89, N 2.88.

{2-[3,7-Di(thiophen-3-yl)-10*H*-phenothiazin-10-yl]ethyl}phosphonic acid (Th-2EPT or V1495 SAM)

Diethyl {2-[3,7-di(thiophen-3-yl)-10*H*-phenothiazin-10-yl]ethyl}phosphonate (**4**, 1 g, 1.89 mmol) was dissolved in anhydrous 1,4-dioxane (25 ml) under argon atmosphere. Afterwards, bromotrimethylsilane (2.5 ml, 18.9 mmol) was added dropwise and reaction was stirred overnight at 25 °C. After consumption of phosphonate **5** (TLC, eluent acetone:n-hexane, 8:17) methanol (0.8 ml, 18.9 mmol) was added and stirring continued for 2 hours. Afterwards, distilled water was added dropwise until precipitate was formed and stirring continued overnight. Product was purified by dissolving in minimum amount of THF, precipitating into 20-fold excess of *n*-hexane, filtering, and washing with *n*-hexane to give brown crystals (0.62 g, 70 % yield). M.p. 198–200 °C (melting and decomposition). $^{1}$H NMR (400 MHz, DMSO-$d_6$): δ 7.80 (s, 2H), 7.62 – 7.57 (m, 2H), 7.58 – 7.47 (m, 6H), 7.01 (d, $J$ = 8.4 Hz, 2H), 4.13 – 3.98 (m, 2H), 2.16 – 2.00 (m, 2H) ppm. $^{13}$C NMR (101 MHz, DMSO-$d_6$): δ 142.33, 140.16, 129.89, 126.98, 125.96, 125.42, 124.45, 122.75, 119.98, 115.17, 41.98, 26.33, 25.03 ppm. Anal. calcd. for $C_{22}H_{18}NO_3PS_3$: C 56.04, H, 3.85, N, 2.97; found: C 56.20, H 4.01, N 2.83.

# NMR spectra of synthesized compounds

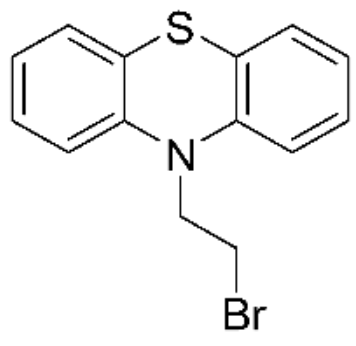

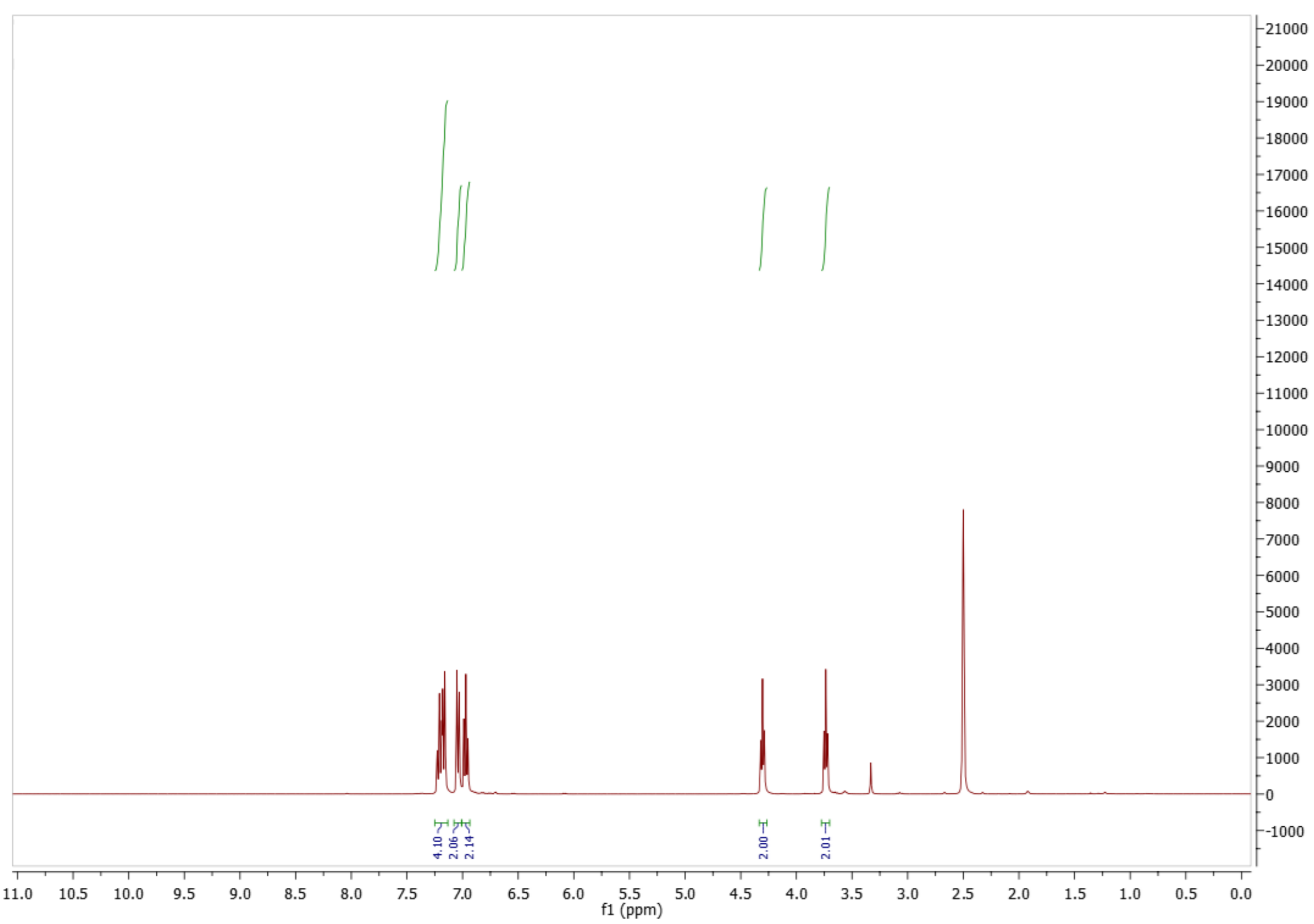

**Figure S11** 6 $^{1}$H NMR spectrum of compound **1** in DMSO-$d_6$

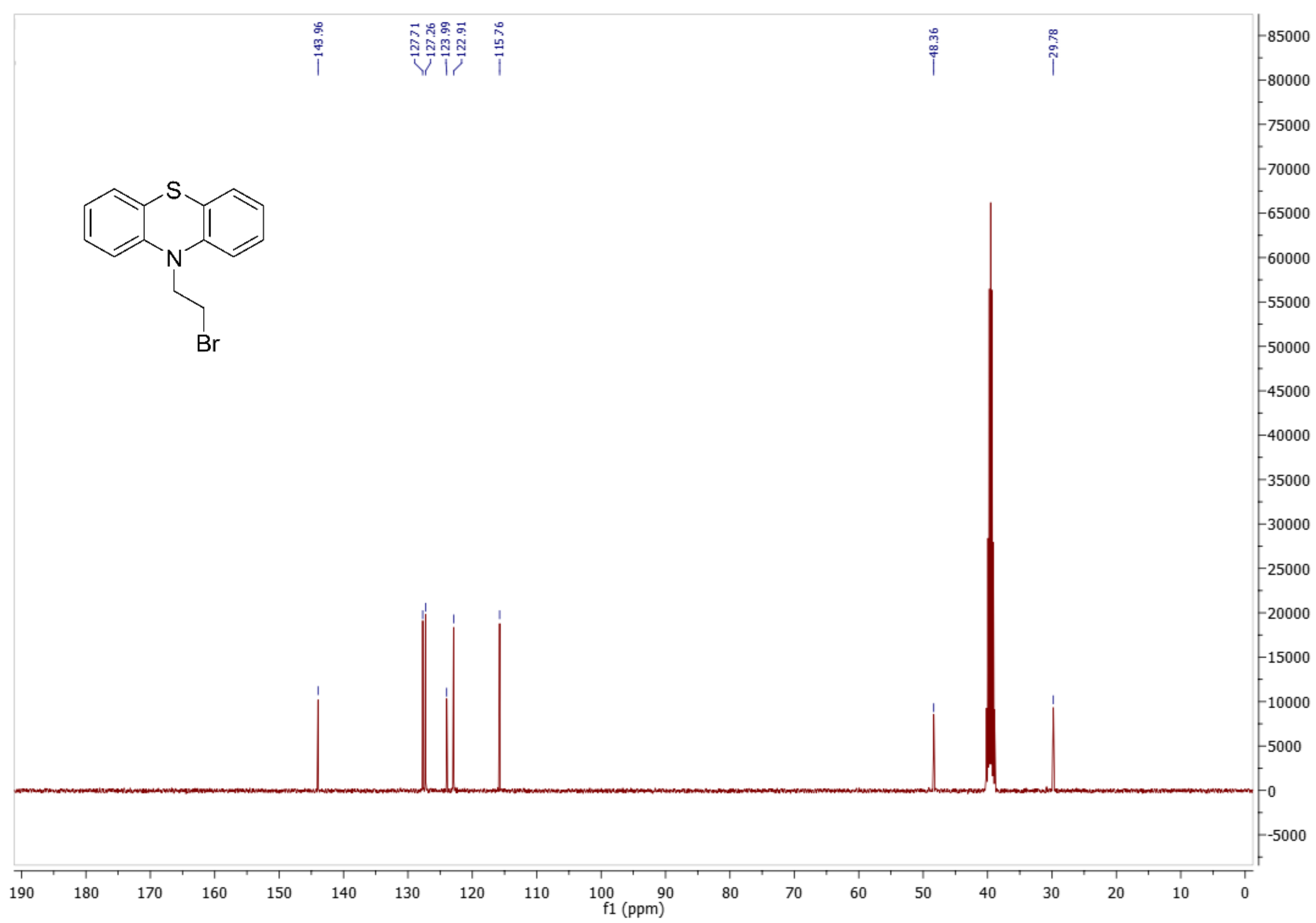

**Figure S12** 7 $^{13}$C NMR spectrum of compound 1 in DMSO-d6

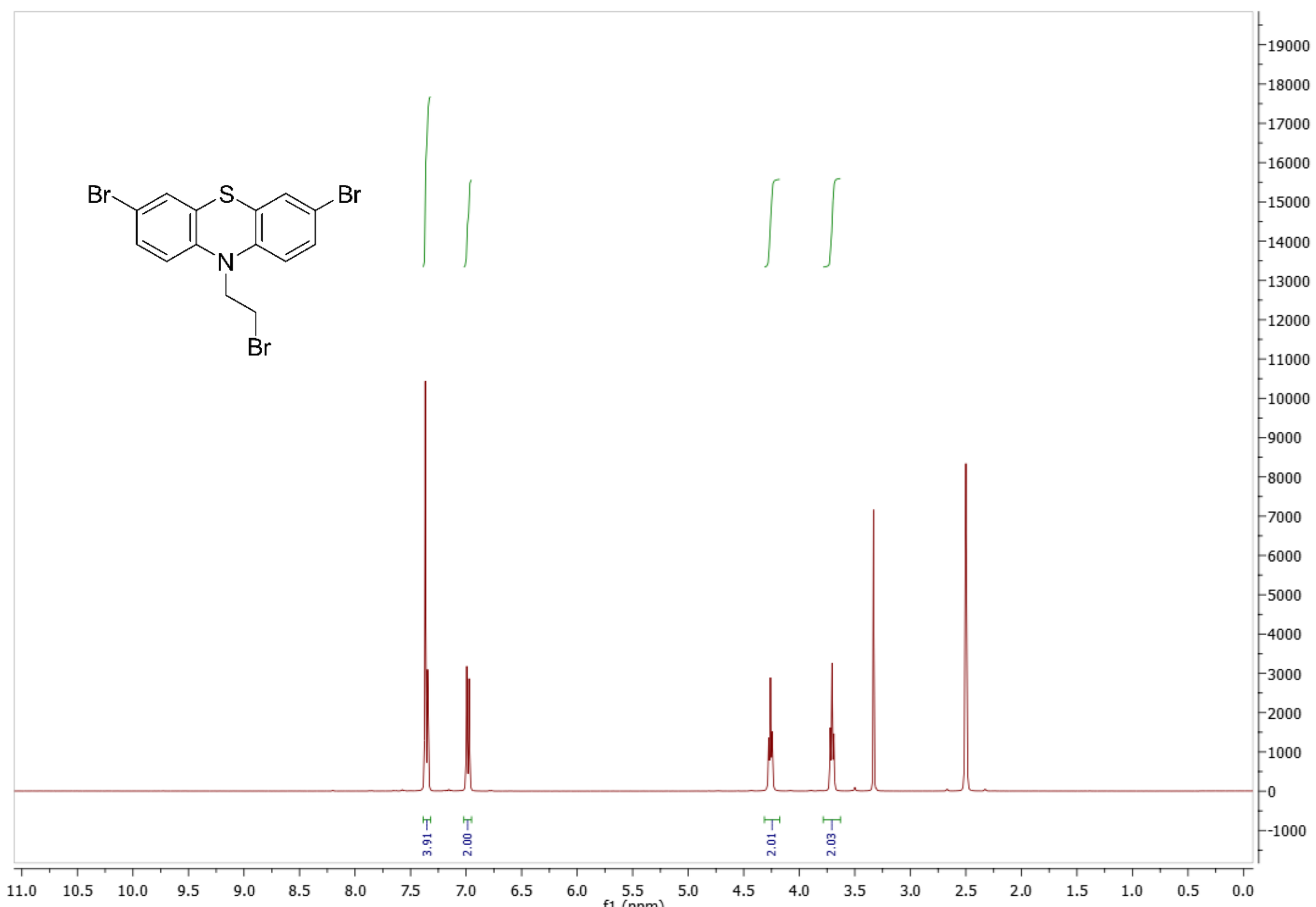

**Figure S13**8 $^1$H NMR spectrum of compound 2 in DMSO-d6

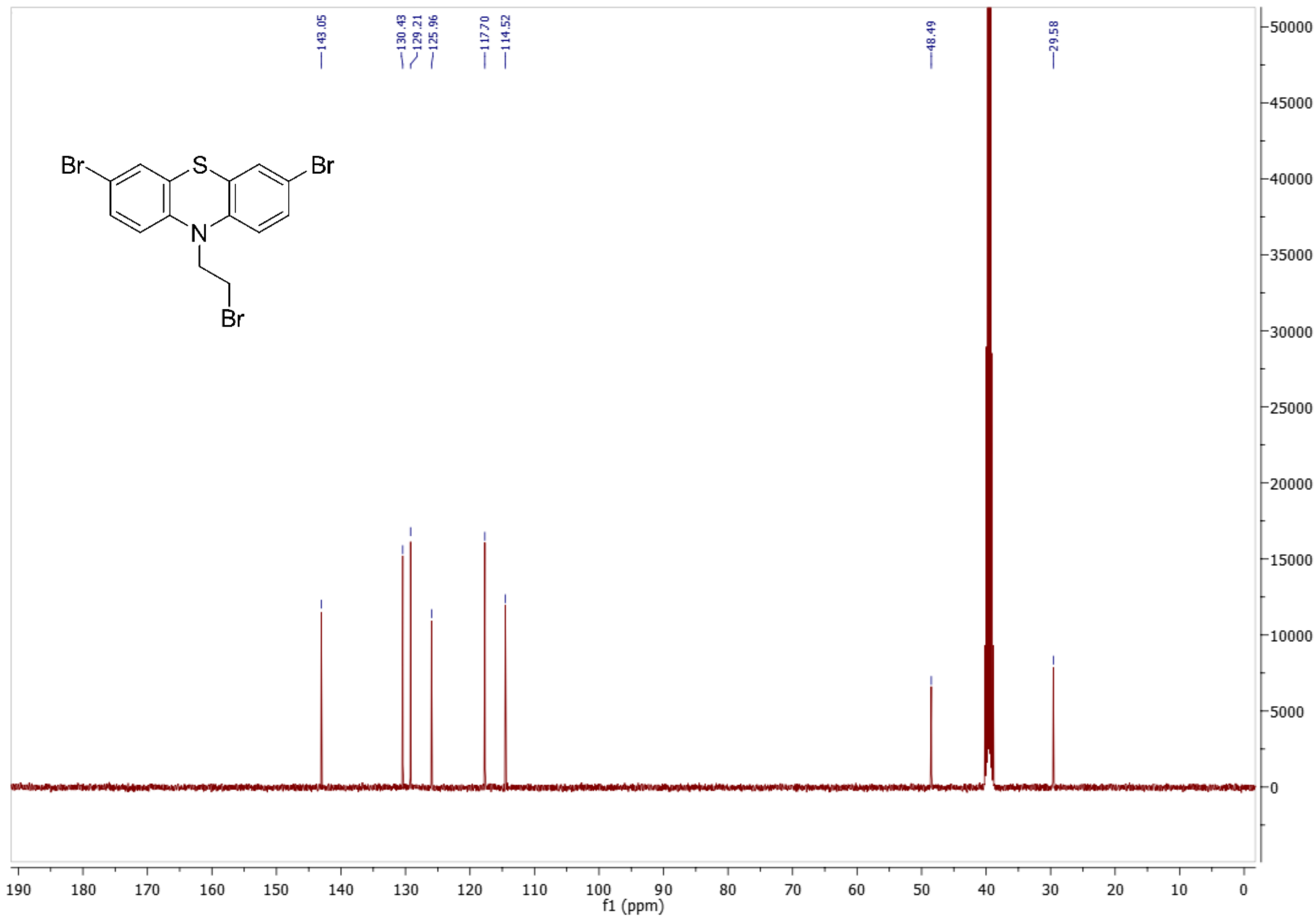

**Figure S14**9 $^{13}$C NMR spectrum of compound 2 in DMSO-d6

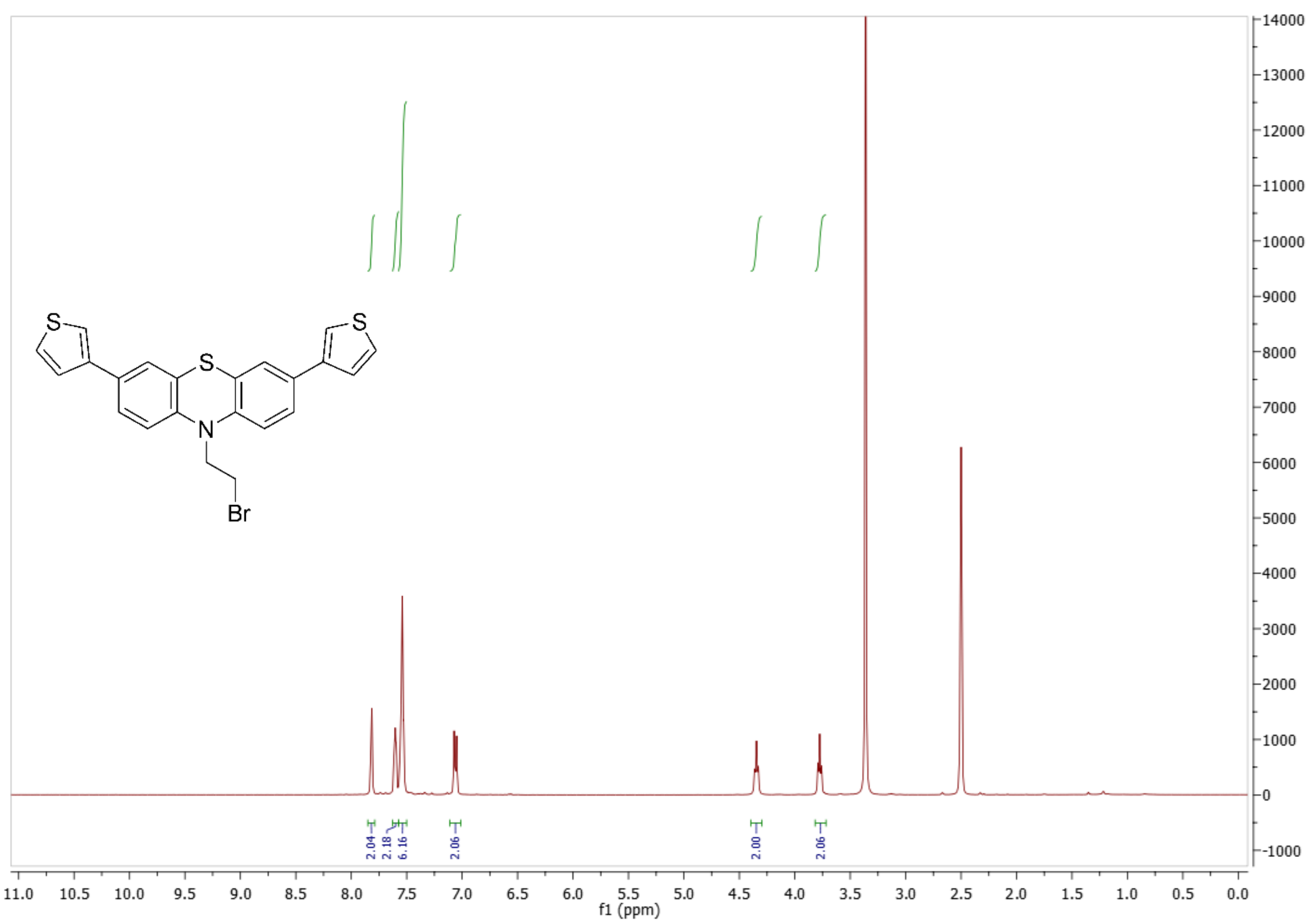

**Figure S1510** $^{1}$H NMR spectrum of compound **3** in DMSO-$d_6$

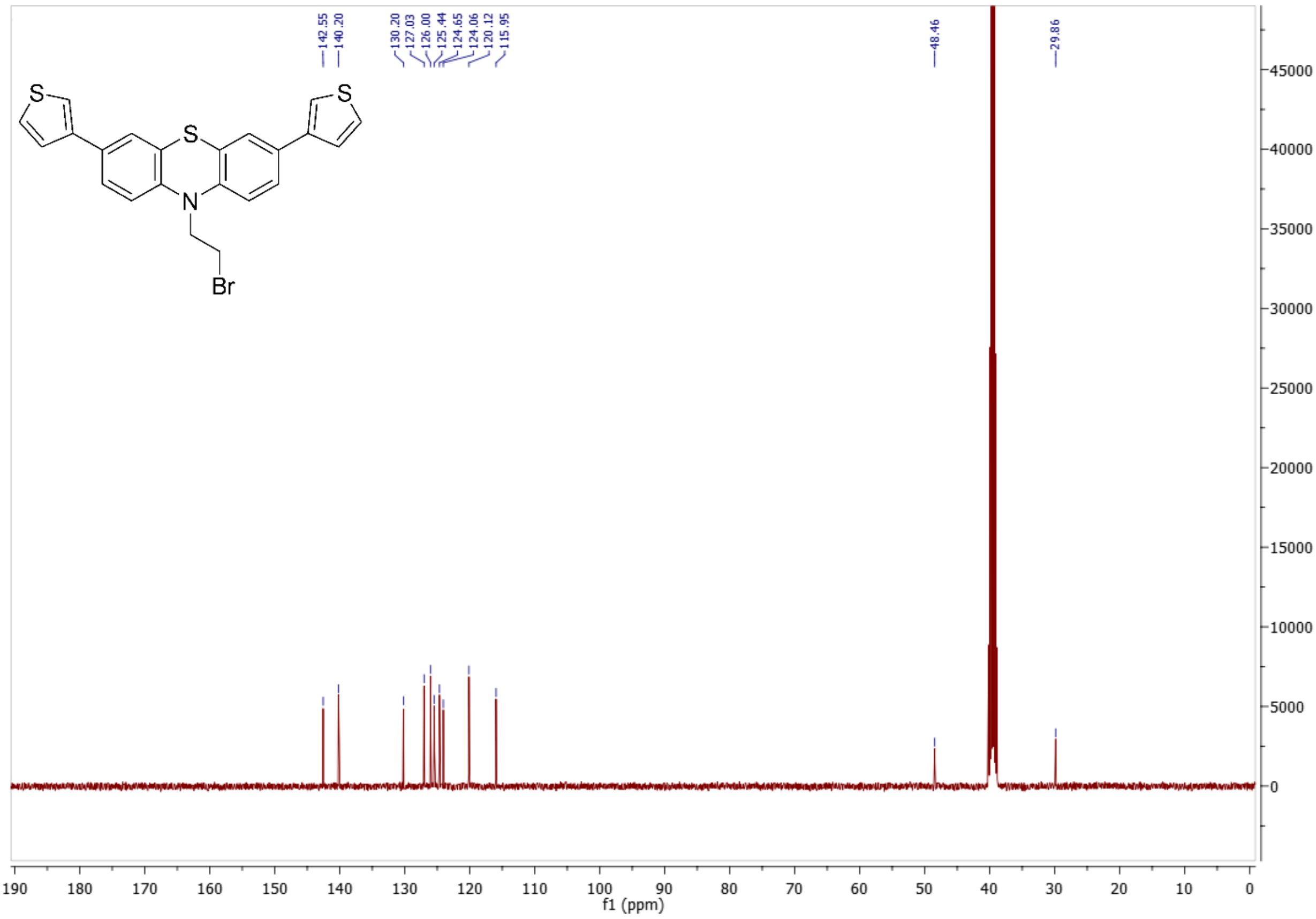

**Figure S16**11 $^{13}$C NMR spectrum of compound **3** in DMSO-$d_6$

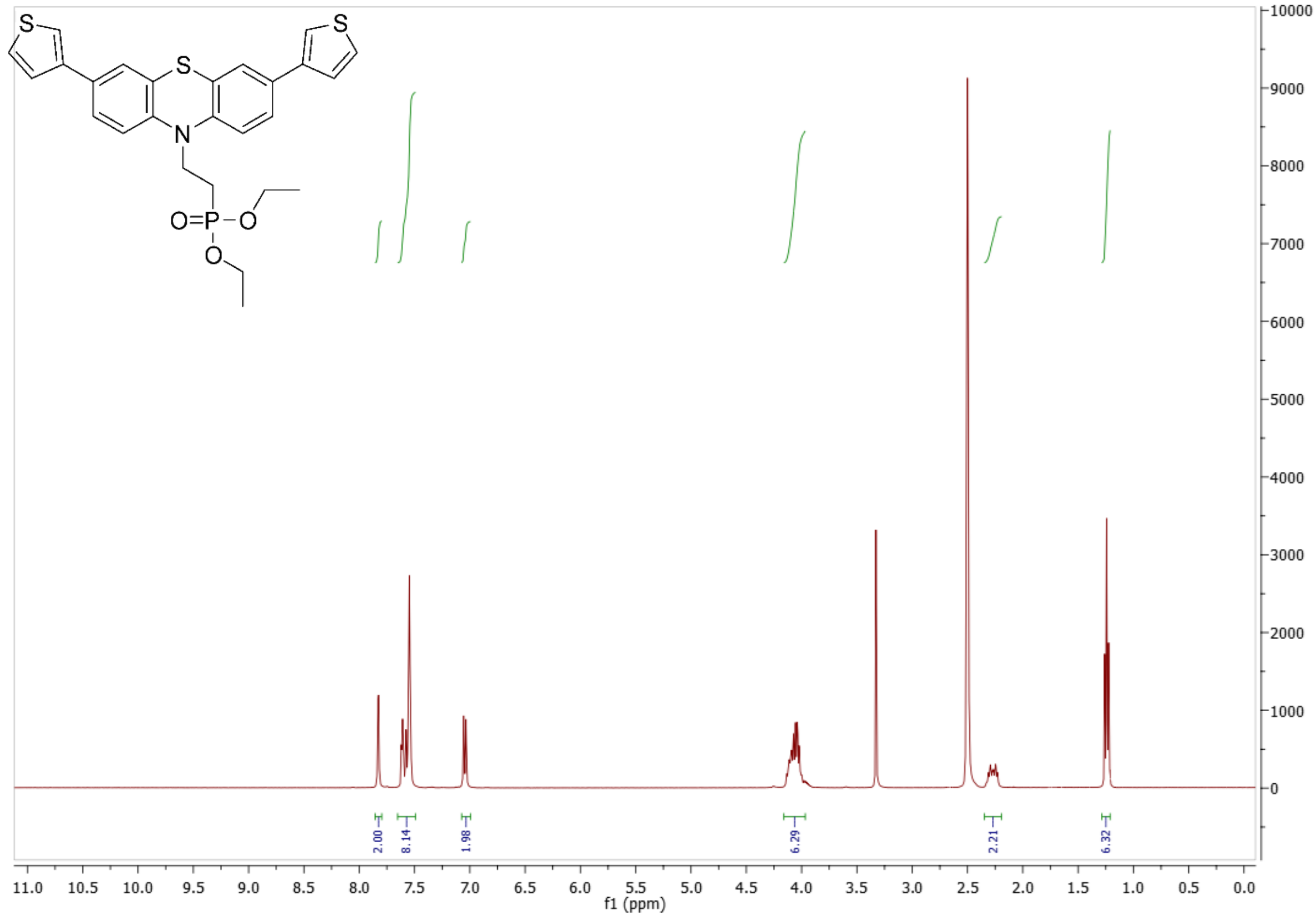

**Figure S17**12 $^{1}$H NMR spectrum of compound **4** in DMSO-$d_6$

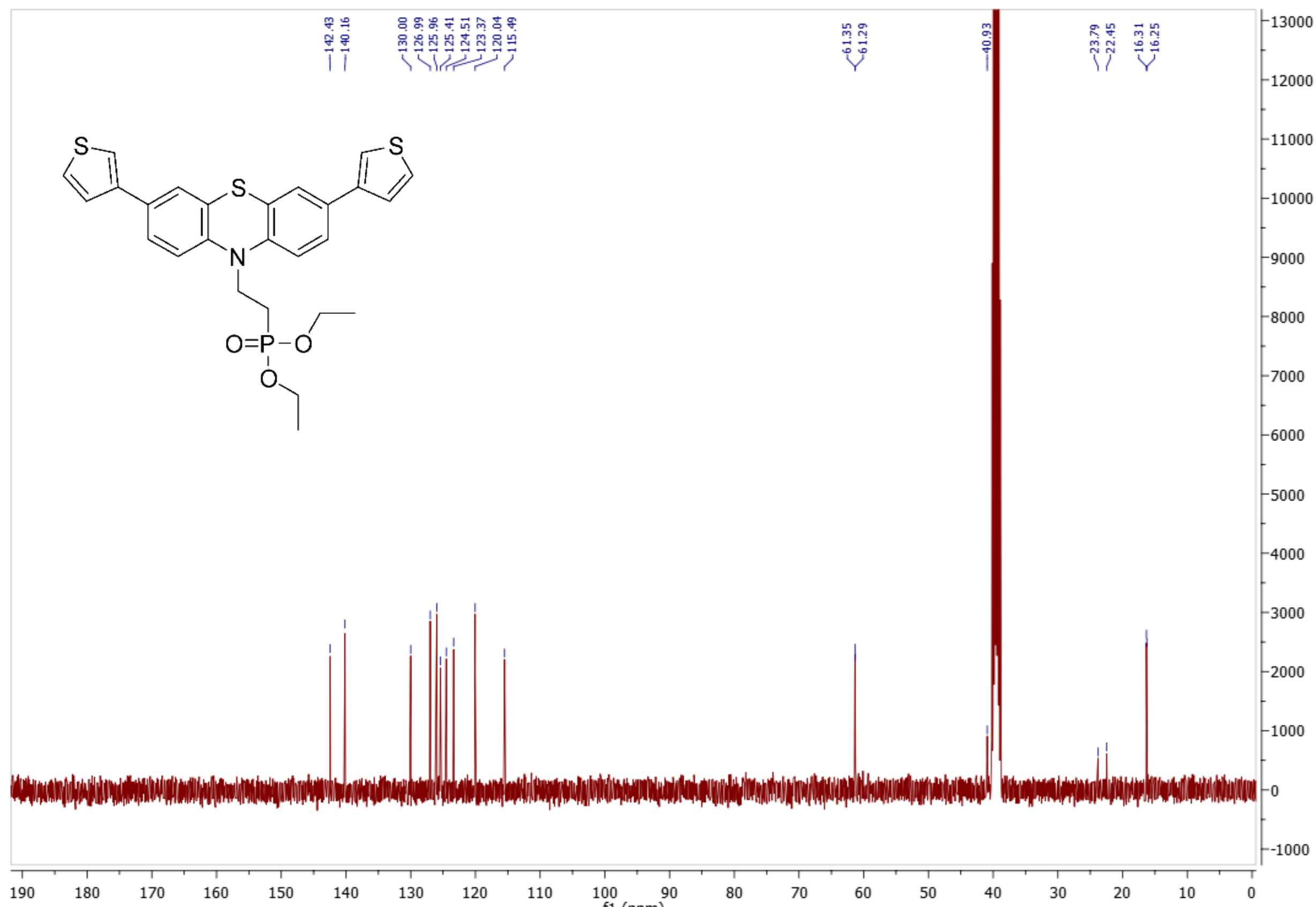

**Figure S18**13 $^{13}$C NMR spectrum of compound **4** in DMSO-$d_6$

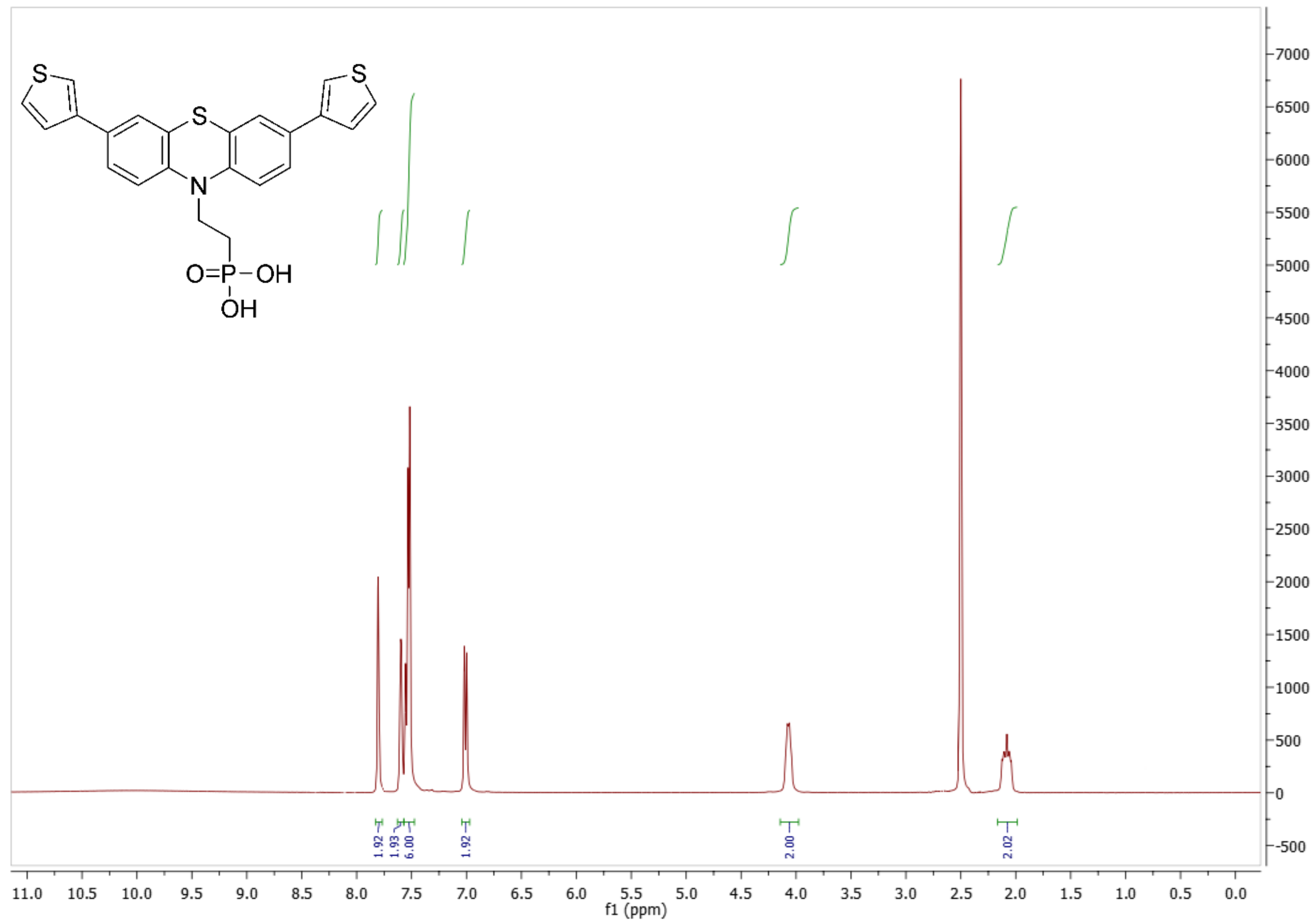

**Figure S19**14 $^{1}$H NMR spectrum of Th-2EPT in DMSO-$d_6$

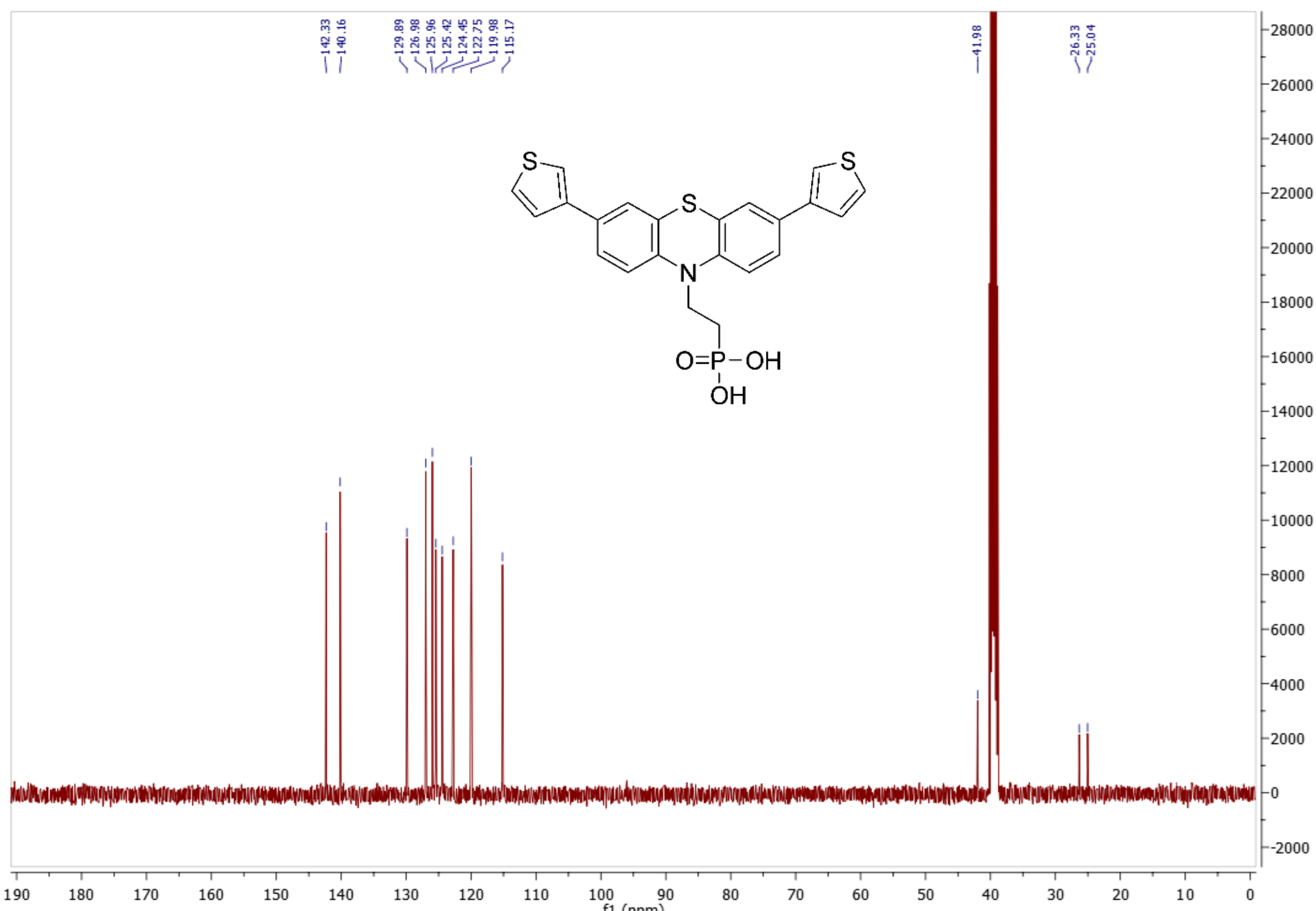

**Figure S20**15 $^{13}$C NMR spectrum of Th-2EPT in DMSO-$d_6$

*Photoelectrical measurements*

The solid-state ionization potential ($I_p$) of the layer of the synthesized Th-2EPT SAM was measured by the electron photoemission in the air method. The sample for the ionization energy measurement was prepared by dissolving material in THF, and the solutions was coated on Al plates that were pre-coated with a ~0.5 μm thick methylmethacrylate and methacrylic acid copolymer adhesive layer. The thickness of the transporting material layer was 0.5–1 μm. The samples were illuminated with monochromatic light from the quartz monochromator with a deuterium lamp. The power of the incident light beam was (2–5)·10-8 W. The negative voltage of –300 V was supplied to the sample substrate. The counter electrode with the 4.5×15 $mm^2$ slit for illumination was placed at an 8 mm distance from the sample surface. The counter-electrode was connected to the input of the BK2-16 type electrometer working in the open input regime for the photocurrent measurement. The 10-15 – 10-12 A strong photocurrent was flowing in the circuit under illumination. Photocurrent I is strongly dependent on the incident light photon energy hν. The $I^{0.5}$ =f(hν) dependence chart was plotted. The linear part of this dependence was extrapolated to the hν axis, and the $I_p$ value was determined as the photon energy at the interception point.

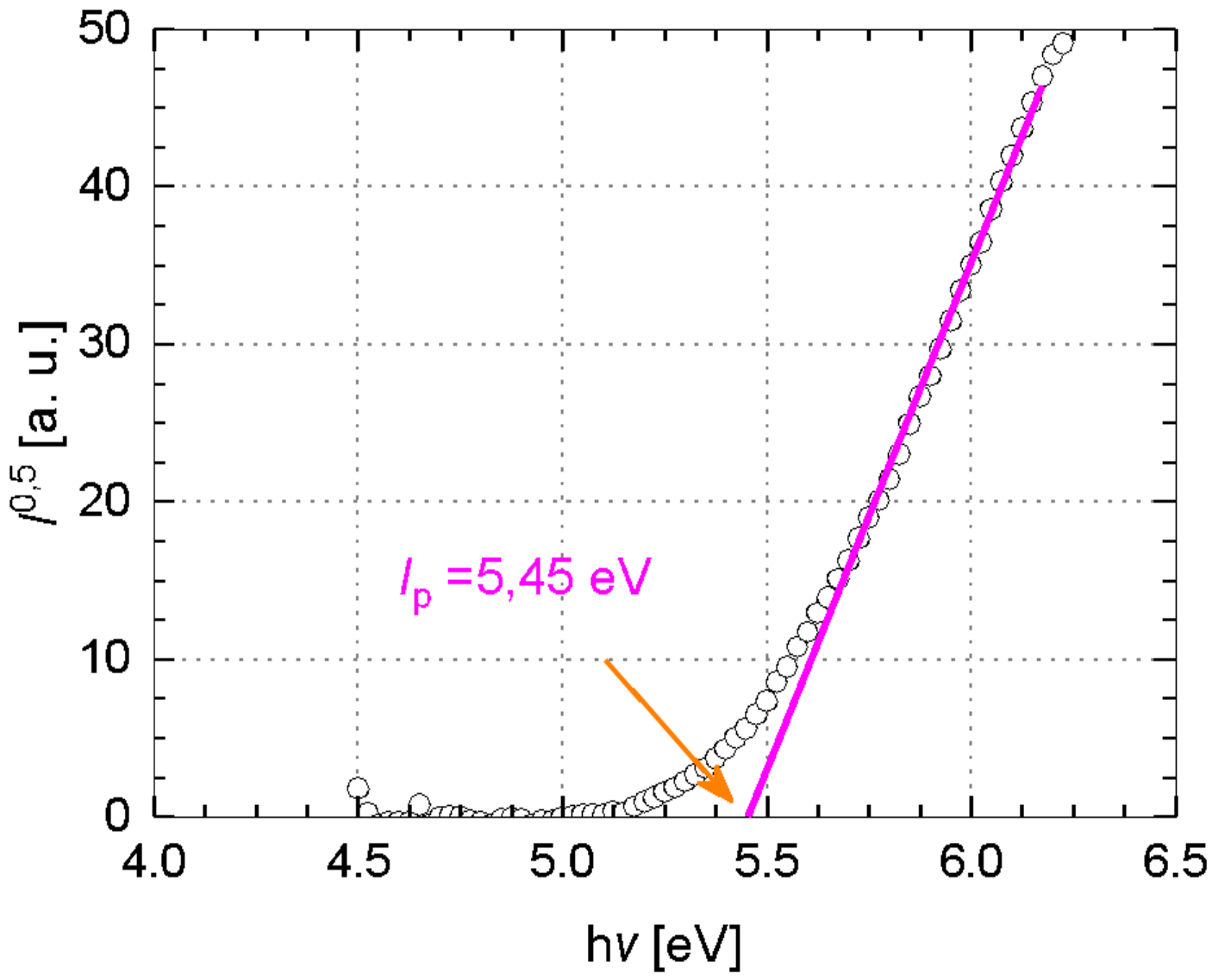

**Figure S21**16 Ip measurement on Th-2EPT

# References

1. Blum, V. *et al.* Ab initio molecular simulations with numeric atom-centered orbitals. *Comput. Phys. Commun.* **180**, 2175–2196 (2009).

2. Stephens, P. J., Devlin, F. J., Chabalowski, C. F. & Frisch, M. J. Ab Initio Calculation of Vibrational Absorption and Circular Dichroism Spectra Using Density Functional Force Fields. *J. Phys. Chem.* **98**, 11623–11627 (1994).

3. Simon, S., Duran, M. & Dannenberg, J. J. Effect of Basis Set Superposition Error on the Water Dimer Surface Calculated at Hartree−Fock, Møller−Plesset, and Density Functional Theory Levels. *J. Phys. Chem. A* **103**, 1640–1643 (1999).

4. Lange, I. *et al.* Tuning the Work Function of Polar Zinc Oxide Surfaces using Modified Phosphonic Acid Self-Assembled Monolayers. *Adv. Funct. Mater.* **24**, 7014–7024 (2014).

5. Liu, X., Wang, L. & Tong, Y. Optoelectronic Properties of Ultrathin Indium Tin Oxide Films: A First-Principle Study. *Crystals* **11**, 30 (2020).

6. Ricciarelli, D., Meggiolaro, D., Ambrosio, F. & De Angelis, F. Instability of Tin Iodide Perovskites: Bulk p-Doping versus Surface Tin Oxidation. *ACS Energy Lett.* **5**, 2787–2795 (2020).